\documentclass[reqno,11pt]{amsart}
\usepackage[utf8]{inputenc}
\usepackage{enumitem}
\usepackage{graphicx}
\usepackage{amscd}
\usepackage{slashed}
\usepackage{amssymb}
\usepackage{mathtools} 
\usepackage{pstricks}
\usepackage[mathscr]{eucal}
\numberwithin{equation}{section}
\allowdisplaybreaks[1]

\title[Causal Fermion Systems]
{Causal Fermion Systems: Discrete Space-Times, Causation and Finite Propagation Speed}

\author[F.\ Finster]{Felix Finster \\ \\ December 2018}
\address{Fakult\"at f\"ur Mathematik \\ Universit\"at Regensburg \\ D-93040 Regensburg \\ Germany}
\email{finster@ur.de}

\newtheorem{Def}{Definition}[section]

\newtheorem{Thm}[Def]{Theorem}

\newcommand{\Thanks}{\vspace*{.5em} \noindent \thanks}
\newcommand{\beq}{\begin{equation}}
\newcommand{\eeq}{\end{equation}}
\newcommand{\Proof}{\begin{proof}}
\newcommand{\QED}{\end{proof} \noindent}

\newcommand{\la}{\langle}
\newcommand{\ra}{\rangle}

\newcommand{\Sl}{\mbox{$\prec \!\!$ \nolinebreak}}
\newcommand{\Sr}{\mbox{\nolinebreak $\succ$}}

\newcommand{\C}{\mathbb{C}}
\newcommand{\R}{\mathbb{R}}
\newcommand{\1}{\mbox{\rm 1 \hspace{-1.05 em} 1}}

\newcommand{\N}{\mathbb{N}}

\DeclareMathOperator{\Tr}{Tr}
\DeclareMathOperator{\tr}{tr}

\renewcommand{\L}{{\mathcal{L}}}
\newcommand{\Sact}{{\mathcal{S}}}

\DeclareMathOperator{\supp}{supp}
\renewcommand{\H}{\mathscr{H}}

\newcommand{\Lin}{\text{\rm{L}}}
\newcommand{\F}{{\mathscr{F}}}

\newcommand{\D}{\mathscr{D}}

\newcommand{\scrM}{\mycal M}

\newcommand{\itemD}{\item[{\raisebox{0.125em}{\tiny $\blacktriangleright$}}]}

\newcommand{\bitem}{\begin{itemize}[leftmargin=2em]}
\newcommand{\eitem}{\end{itemize}}
\renewcommand{\u}{\mathfrak{u}}
\renewcommand{\v}{\mathfrak{v}}

\newcommand{\h}{\mathfrak{h}}

\DeclareFontFamily{OT1}{rsfso}{}
\DeclareFontShape{OT1}{rsfso}{m}{n}{ <-7> rsfso5 <7-10> rsfso7 <10-> rsfso10}{}
\DeclareMathAlphabet{\mycal}{OT1}{rsfso}{m}{n}

\begin{document}

\maketitle

\begin{abstract}
The theory of causal fermion systems is a recent approach to fundamental physics. Giving quantum mechanics, general relativity and quantum field theory as limiting cases, it is a candidate for a unified physical theory. The dynamics is described by a novel variational principle, the so-called causal action principle. The causal action principle does not rely on a presupposed space-time structure. Instead, it is a variational principle for space-time itself as well as for all structures in space-time (like particles, fields, etc.).

After a general motivation and introduction, we report on mathematical results for
two-particle causal fermion systems which state that
every minimizer describes a discrete space-time. We explain and make precise that on scales which are much
larger than the scale of the microscopic space-time structures, the dynamics of a causal fermion system
respects causality with a finite speed of propagation.
\end{abstract}
\tableofcontents

\section{From a Lattice System to Causal Fermion Systems}
The focus of this conference is on discrete space-time structures.
Already in previous talks the motivation for thinking about discrete space-times
was mentioned: The ultraviolet divergences of quantum field
theory indicate that the structure of space-time should be modified on a microscopic scale,
and the Planck length gives a natural length scale for such modifications and for new physics.
To my opinion, instead of taking sides in favor of discrete or continuous microstructures,
it is more promising to consider the following more general question:
\begin{quote}
What is the structure of space-time on the Planck scale?
\end{quote}
I shall present a concise proposal for what the mathematical structure of microscopic space-time
could be. The mathematical structures are quite general and allow for the description of
both discrete and continuum space-times. Nevertheless, as we shall see,
there are results in favor of a discrete microstructure.

\subsection{Example: A Two-Dimensional Lattice System}
\begin{figure}
%
\psscalebox{1.0 1.0} 
{
\begin{pspicture}(0,-1.6720673)(7.0918274,1.6720673)
\definecolor{colour0}{rgb}{0.0,0.6,0.4}
\definecolor{colour1}{rgb}{0.8,0.2,0.0}
\definecolor{colour2}{rgb}{0.0,0.4,0.0}
\psdots[linecolor=colour0, dotsize=0.12](1.105,-0.78706735)
\psdots[linecolor=colour1, dotsize=0.12](1.105,0.81293267)
\psdots[linecolor=black, dotsize=0.12](1.105,1.6129327)
\psdots[linecolor=black, dotsize=0.12](1.905,1.6129327)
\psdots[linecolor=black, dotsize=0.12](2.705,1.6129327)
\psdots[linecolor=black, dotsize=0.12](3.505,1.6129327)
\psdots[linecolor=black, dotsize=0.12](4.305,1.6129327)
\psdots[linecolor=black, dotsize=0.12](5.105,1.6129327)
\rput[bl](5.375,-0.90206736){\normalsize{$t-\Delta t$}}
\rput[bl](5.385,0.6729327){\normalsize{$t+\Delta t$}}
\rput[bl](5.665,-0.087067336){\normalsize{$t$}}
\psdots[linecolor=colour0, dotsize=0.12](1.905,-0.78706735)
\psdots[linecolor=colour0, dotsize=0.12](2.705,-0.78706735)
\psdots[linecolor=colour0, dotsize=0.12](3.505,-0.78706735)
\psdots[linecolor=colour0, dotsize=0.12](4.305,-0.78706735)
\psdots[linecolor=colour0, dotsize=0.12](5.105,-0.78706735)
\psdots[linecolor=colour0, dotsize=0.12](1.105,0.012932663)
\psdots[linecolor=colour0, dotsize=0.12](1.905,0.012932663)
\psdots[linecolor=colour0, dotsize=0.12](2.705,0.012932663)
\psdots[linecolor=colour0, dotsize=0.12](3.505,0.012932663)
\psdots[linecolor=colour0, dotsize=0.12](4.305,0.012932663)
\psdots[linecolor=colour0, dotsize=0.12](5.105,0.012932663)
\psdots[linecolor=colour1, dotsize=0.12](1.905,0.81293267)
\psdots[linecolor=colour1, dotsize=0.12](2.705,0.81293267)
\psdots[linecolor=colour1, dotsize=0.12](3.505,0.81293267)
\psdots[linecolor=colour1, dotsize=0.12](4.305,0.81293267)
\psdots[linecolor=colour1, dotsize=0.12](5.105,0.81293267)
\psbezier[linecolor=colour1, linewidth=0.04, arrowsize=0.05291667cm 2.0,arrowlength=1.4,arrowinset=0.0]{->}(6.895,0.057932664)(7.18,0.36793268)(7.03,0.53293264)(6.9,0.7579326629638672)
\psbezier[linecolor=colour1, linewidth=0.04, arrowsize=0.05291667cm 2.0,arrowlength=1.4,arrowinset=0.0]{->}(6.89,0.90793264)(7.175,1.2179327)(7.025,1.3829327)(6.895,1.6079326629638673)
\rput[bl](2.405,-0.30706733){\normalsize{$+$}}
\rput[bl](4.395,-0.30206734){\normalsize{$+$}}
\rput[bl](3.39,-1.0970674){\normalsize{$-$}}
\pspolygon[linecolor=colour2, linewidth=0.02](2.7089655,0.014901032)(3.5057106,-0.7929771)(4.3039584,0.008308901)
\psline[linecolor=colour1, linewidth=0.02, arrowsize=0.05291667cm 2.0,arrowlength=1.4,arrowinset=0.0]{->}(3.495,0.22793266)(3.5,0.6029327)
\psline[linecolor=black, linewidth=0.04](1.095,-1.0920674)(1.195,-1.1920674)(1.395,-1.1920674)(1.495,-1.2920673)(1.595,-1.1920674)(1.795,-1.1920674)(1.895,-1.0920674)
\psline[linecolor=black, linewidth=0.04](0.79,-0.78206736)(0.69,-0.68206733)(0.69,-0.48206735)(0.59,-0.38206732)(0.69,-0.28206733)(0.69,-0.08206734)(0.79,0.017932663)
\rput[bl](0.0,-0.48206735){\normalsize{$\Delta t$}}
\rput[bl](1.215,-1.6720673){\normalsize{$\Delta x$}}
\end{pspicture}
}
\caption{Time evolution of a lattice system~$\scrM \subset \R^{1,1}$.}
\label{lattice2}
\end{figure}
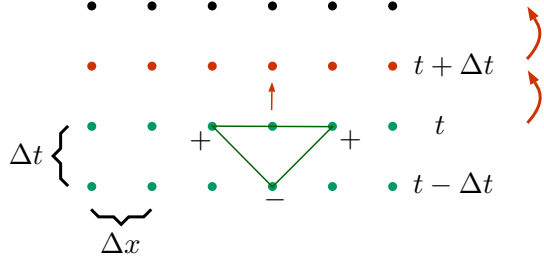

For the motivation, I would like to begin with the familiar example of a space-time lattice.
In order to keep the setting as simple as possible, we consider a two-dimensional lattice
(one space and one time dimension), but higher-dimensional lattices can be described similarly.
Thus let~$\scrM \subset \R^{1,1}$ be a lattice in two-dimensional Minkowski space.
We denote the spacing in time direction by~$\Delta t$ and in spatial direction by~$\Delta x$
(see Figure~\ref{lattice2}).
The usual procedure for setting up equations on a lattice is to replace derivatives
by difference quotients, giving rise to an evolution equation which can be
solved time step by time step according to deterministic rules.

As a concrete example, let us consider a discretization of the
two-dimensional wave equation on the lattice,
\beq \label{wavelattice}
\begin{split}
0 = \Box \phi(t,x) &:= \frac{1}{(\Delta t)^2} \Big( \phi(t+\Delta t, x) 
- 2 \phi(t, x) + \phi(t-\Delta t, x)\Big) \\
&\quad\; - \frac{1}{(\Delta x)^2} \Big( \phi(t, x+\Delta x) 
- 2 \phi(t, x) + \phi(t, x - \Delta x)\Big) \:.
\end{split}
\eeq
Solving this equation for~$\phi(t+\Delta t, x)$ gives a rule
for computing~$\phi(t+\Delta t, x)$ from the values of~$\phi$
at earlier times~$t$ and~$t-\Delta t$ (see again Figure~\ref{lattice2}).

While this method is very simple and gives well-defined evolution equations, it
also has several drawbacks:
\bitem
\itemD The above method of discretizing the continuum equations
is very {\em{ad hoc}}. Why do we choose a regular lattice, why do we work with difference quotients?
There are many other ways of discretizing the wave equation.
\itemD The method is {\em{not background-free}}. In order to speak of the ``lattice spacing,''
the lattice must be thought of as being embedded in a two-dimensional ambient space-time.
\itemD The concept of a space-time lattice is not invariant under general coordinate transformations.
In other words, the assumption of a space-time lattice is
{\em{not compatible with the equivalence principle}}.
\eitem
In view of these shortcomings, the following basic question arises:
\begin{quote}
Can one formulate equations without referring
to the nearest neighbor relation and the lattice spacing?
\end{quote}
The answer to this question is yes, 
and we will now see how this can be done in the example of our two-dimensional lattice system.
Although our example is somewhat oversimplified, this consideration will lead us quite naturally to the setting
of causal fermion systems.

We consider complex-valued wave functions~$\psi_1, \ldots, \psi_f$
on the lattice~$\scrM$ (for simplicity a finite number, i.e.\ $f < \infty$).
These wave functions do not need to satisfy the wave equation.
On the complex vector space spanned by these wave functions we introduce
a scalar product~$\la .|. \ra_\H$ and assume that the wave functions are orthonormal, i.e.\
\beq \label{ortho}
\la \psi_k | \psi_l \ra_\H = \delta_{kl} \:.
\eeq
We thus obtain an $f$-dimensional Hilbert space~$(\H, \la .|. \ra_\H)$.
Note that the scalar product is given abstractly (meaning that it has no
representation in terms of the wave functions as a sum over lattice points).
Next, for any lattice point~$(t,x) \in \scrM$ we introduce the
so-called {\em{local correlation operator}}~$F(t,x) : \H \rightarrow \H$
as a matrix with elements given by
\beq \label{matrix}
(F(t,x))^j_k = \overline{\psi_j(t,x)} \psi_k(t,x) \:.
\eeq
The diagonal elements of this matrix are the absolute squares~$|\psi_k(t,x)|^2$
of the corresponding wave functions. The off-diagonal elements, on the other hand,
tell us about the correlation of the $j^\text{th}$ and $k^\text{th}$ wave functions
at the lattice point~$(t,x)$. This is the reason for the name ``local correlation operator.''
This operator can also be characterized in a basis-invariant way by the relations
\[ \la \psi, F(t,x) \,\phi \ra_\H = \overline{\psi(t,x)} \phi(t,x) \:, \]
to be satisfied for all~$\psi, \phi \in \H$.
Taking the complex conjugate, one sees immediately that the matrix defined
by~\eqref{matrix} is Hermitian. Stated equivalently independent of bases,
the local correlation operator is a {\em{selfadjoint}} linear operator on~$\H$. Moreover, a local correlation operator
has {\em{rank at most one}} and is {\em{positive semi-definite}}. This can be seen by writing it as
\beq \label{ees}
F(t,x) = e^* e \qquad \text{with} \qquad e : \H \rightarrow \C\:,\quad
\psi \mapsto \psi(x) \:.
\eeq

It is useful to denote the set of all operators with the above properties by~$\F$,
\[ \F := \big\{ F \in \Lin(\H) \:|\: \text{$F$ is selfadjoint, positive semi-definite and has rank at most one} \} \:. \]
Varying the lattice point, we obtain a mapping (see Figure~\ref{embed})
\[ F : \scrM \rightarrow \F \:,\qquad (t,x) \mapsto F(t,x) \:. \]
\begin{figure}
%
\psscalebox{1.0 1.0} 
{
\begin{pspicture}(-1.5,-1.7327906)(10.322246,1.7327906)
\definecolor{colour0}{rgb}{0.0,0.0,0.6}
\pspolygon[linecolor=white, linewidth=0.02, fillstyle=solid](1.1613568,0.65876496)(1.2102457,0.52543163)(1.3169124,0.42765382)(1.4813569,0.40098715)(1.6502457,0.40987605)(1.743579,0.4632094)(1.5791346,0.5832094)(1.3613569,0.65876496)
\rput[bl](5.597801,-1.7327906){$0$}
\rput[bl](6.562246,-0.23501284){$\F \subset \Lin(\H)$}
\pspolygon[linecolor=white, linewidth=0.02, fillstyle=solid,fillcolor=yellow](3.6458013,1.7209872)(5.479579,-1.5207906)(7.2120233,1.722765)
\psline[linecolor=black, linewidth=0.04](3.663579,1.6898761)(5.472468,-1.5385684)(7.201357,1.7103205)
\rput[bl](0.4880235,0.9576538){$\scrM$}
\psbezier[linecolor=colour0, linewidth=0.04, arrowsize=0.05291667cm 2.0,arrowlength=1.4,arrowinset=0.0]{->}(2.4033267,0.116877794)(2.6445446,-0.122063294)(3.155987,-0.51524955)(4.3238316,-0.02157013655146784)
\rput[bl](3.2280235,0.03876494){\textcolor{colour0}{$F$}}
\pspolygon[linecolor=yellow, linewidth=0.02, fillstyle=solid,fillcolor=yellow](5.050246,0.51400304)(5.084468,0.4511459)(5.1581774,0.39114588)(5.2397842,0.3654316)(5.3398185,0.3711459)(5.392468,0.40257445)(5.2950664,0.4797173)(5.163442,0.52543163)
\psdots[linecolor=black, dotsize=0.12](1.6591346,0.6632094)
\psdots[linecolor=black, dotsize=0.12](0.0591346,0.6632094)
\psdots[linecolor=black, dotsize=0.12](0.4591346,0.6632094)
\psdots[linecolor=black, dotsize=0.12](0.8591346,0.6632094)
\psdots[linecolor=black, dotsize=0.12](1.2591347,0.6632094)
\psdots[linecolor=black, dotsize=0.12](1.6591346,0.26320937)
\psdots[linecolor=black, dotsize=0.12](0.0591346,0.26320937)
\psdots[linecolor=black, dotsize=0.12](0.4591346,0.26320937)
\psdots[linecolor=black, dotsize=0.12](0.8591346,0.26320937)
\psdots[linecolor=black, dotsize=0.12](1.2591347,0.26320937)
\psdots[linecolor=black, dotsize=0.12](1.6591346,-0.13679062)
\psdots[linecolor=black, dotsize=0.12](0.0591346,-0.13679062)
\psdots[linecolor=black, dotsize=0.12](0.4591346,-0.13679062)
\psdots[linecolor=black, dotsize=0.12](0.8591346,-0.13679062)
\psdots[linecolor=black, dotsize=0.12](1.2591347,-0.13679062)
\psdots[linecolor=black, dotsize=0.12](1.6591346,-0.5367906)
\psdots[linecolor=black, dotsize=0.12](0.0591346,-0.5367906)
\psdots[linecolor=black, dotsize=0.12](0.4591346,-0.5367906)
\psdots[linecolor=black, dotsize=0.12](0.8591346,-0.5367906)
\psdots[linecolor=black, dotsize=0.12](1.2591347,-0.5367906)
\psdots[linecolor=black, dotsize=0.08](4.7991347,0.90098715)
\psdots[linecolor=black, dotsize=0.08](5.1391344,0.9309872)
\psdots[linecolor=black, dotsize=0.08](5.4591346,0.9909872)
\psdots[linecolor=black, dotsize=0.08](5.7891345,1.0509871)
\psdots[linecolor=black, dotsize=0.08](6.1591344,1.2509872)
\psdots[linecolor=black, dotsize=0.08](6.1491346,0.96098715)
\psdots[linecolor=black, dotsize=0.08](5.8891344,0.77098715)
\psdots[linecolor=black, dotsize=0.08](5.4991345,0.69098717)
\psdots[linecolor=black, dotsize=0.08](5.1191344,0.64098716)
\psdots[linecolor=black, dotsize=0.08](4.8591347,0.54098713)
\psdots[linecolor=black, dotsize=0.08](5.0791345,0.28098717)
\psdots[linecolor=black, dotsize=0.08](5.2791348,0.35098717)
\psdots[linecolor=black, dotsize=0.08](5.509135,0.43098715)
\psdots[linecolor=black, dotsize=0.08](5.6891346,0.45098716)
\psdots[linecolor=black, dotsize=0.08](5.9291344,0.49098715)
\psdots[linecolor=black, dotsize=0.08](5.1791344,0.030987162)
\psdots[linecolor=black, dotsize=0.08](5.3591347,0.08098716)
\psdots[linecolor=black, dotsize=0.08](5.6091347,0.18098716)
\psdots[linecolor=black, dotsize=0.08](5.759135,0.19098715)
\psdots[linecolor=black, dotsize=0.08](6.0291348,0.21098717)
\rput[bl](4.8180237,1.1376538){$F(\scrM)$}
\end{pspicture}
}
\caption{Embedding in~$\F$.}
\label{embed}
\end{figure}%
For clarity, we note that the set~$\F$ is {\em{not}} a vector space,
because the linear combination of operators in~$\F$ will in general have
rank bigger than one. But it is a {\em{conical}} set in the sense
that a positive multiple of any operator in~$\F$ is again in~$\F$
(this is why in Figure~\ref{embed} the set~$\F$ is depicted as a cone).

We point out that the local correlation operators do not involve the lattice spacing
or the nearest neighbor relation; instead they contain information only on
the local correlations of the wave functions at each lattice point.
With this in mind, our strategy for formulating equations which do not involve
the structures of the lattice is to work exclusively with the local correlation operators, i.e.\
with the subset~$F(\scrM) \subset \F$. In other words, in Figure~\ref{embed} we
want to disregard the lattice on the left and work only with the objects on the right.

How can one set up equations purely in terms of the local correlation operators?
In order to explain the general procedure, we consider a finite number of
operators~$F_1, \ldots, F_L$ of~$\F$.
Each of these operators can be thought of as giving information
on the local correlations of the wave functions at a
space-time point. However, this ``space-time point'' is no longer a lattice point,
but at the moment it is merely a point without additional structure.
In order to obtain a ``space-time'' in the usual sense, one needs additional
structures and relations between the space-time points.
Such relations can be obtained by multiplying the operators.
Indeed, the operator product~$F_i \,F_j$ tells us about correlations of the wave functions
at different space-time points. Taking the trace of this operator product gives a
real number. Our method for formulating physical equations is to set up a variational
principle. This variational formulation has the advantage that symmetries give rise to
conservation laws by Noether's theorem (as will be explained in Section~\ref{secosi}).
Therefore, we want to minimize an action~$\Sact$.
A simple example is to
\beq \label{Ssum}
\text{minimize} \qquad \Sact(F_1, \ldots, F_n) := \sum_{i,j=1}^L \Tr(F_i \,F_j)^2
\eeq
under variations of the points~$F_1,\ldots, F_L \in \F$.
In order to obtain a mathematically sensible variational principle,
one needs to impose certain constraints.
Here we do not enter the details, because the present example is a bit too simple.
Instead, we merely use it as a motivation for the general setting of causal fermion
systems, which we now introduce.

\subsection{The Setting of Causal Fermion Systems}
In order to get from our example to the general setting of causal fermion systems,
we must extend the above constructions in several steps:
\bitem
\item[(a)] The previous example works similarly in higher dimensions,
in particular for a lattice~$\scrM \subset \R^{1,3}$ in four-dimensional Minkowski space.
This has no effect on the resulting structure of a finite number of distinguished
operators~$F_1, \ldots, F_L \in \F$.
\item[(b)] Suppose that on the lattice we consider multi-component wave
functions~$\psi(t,x) \in \C^N$. Then the pointwise product on the right side of~\eqref{matrix}
must be replaced by a complex inner product, which we denote by~$\Sl .|. \Sr$
(in mathematical terms, this inner product is a non-degenerate sesquilinear form;
we always use the convention that the wave function on the left is complex conjugated).
Thus the definition of the local correlation operator~\eqref{matrix} is replaced by
\[ 
(F(t,x))^j_k = -\Sl \psi_j(t,x) | \psi_k(t,x) \Sr \]
(the minus sign compared to~\eqref{matrix} merely is a useful convention).
The resulting local correlation operator is no longer an operator of rank
at most one, but it has rank at most~$N$ (as can be seen for example by writing
it similar to~\eqref{ees} in the form~$F(t,x)=-e^* e$ with~$e : \H \rightarrow \C^N$).
If the inner product~$\Sl .|. \Sr$ on~$\C^N$ is positive definite, then the operator~$F(t,x)$
is negative semi-definite.
However, in the physical applications, this inner product will {\em{not}} be
positive definite. Indeed, a typical example in mind is that of four-component Dirac spinors.
The Lorentz invariant inner product~$\overline{\psi} \phi$ on Dirac spinors in Minkowski space
(with the usual adjoint spinor~$\overline{\psi} := \psi^\dagger \gamma^0$)
is indefinite of signature~$(2,2)$.
In order to describe systems involving leptons and quarks, one must take 
direct sums of Dirac spinors, giving the signature~$(n, n)$ with~$n \in 2\N$.
With this in mind, we assume more generally that 
\[ \Sl .|. \Sr \quad \text{has signature~$(n,n)$ with~$n \in \N$}\:. \]
Then the resulting local correlation operators are selfadjoint operators of rank at most~$2n$,
which (counting multiplicities) have at most~$n$ positive and at most~$n$ negative eigenvalues.
\item[(c)] Finally, it is useful to generalize the setting such as to allow for continuous space-times
and for space-times which have both continuous and discrete components. In preparation, we note that
the sums over the operators~$F_1,\ldots, F_L$ in~\eqref{Ssum} can be written as integrals,
\beq \label{Scont}
\Sact(\rho) = \int_\F d\rho(x) \int_\F d\rho(y)\: \Tr(F_i F_j)^2 \:,
\eeq
if~$\rho$ is a measure on~$\F$ chosen as the sum of Dirac measures
supported at these operators,
\beq \label{deltasum}
\rho = \sum_{i=1}^L \delta_{F_i} \:.
\eeq
In this formulation, the measure plays a double role: First, it distinguishes the
points~$F_1, \ldots, F_L$
as those points where the measure is non-zero, as is made mathematically precise by the notion
of the {\em{support}} of the measure, i.e.\
\beq \label{FL}
\text{supp}\, \rho = \{F_1, \ldots, F_L \}\:.
\eeq
Second, a measure makes it possible to integrate over its support, an operation which for
the measure~\eqref{deltasum} reduces to the sum over~$F_1, \ldots, F_L$.

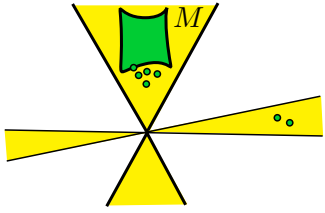
\begin{figure}
{
\begin{pspicture}(0,-1.3610485)(4.2347274,1.3610485)
\definecolor{colour0}{rgb}{0.0,0.8,0.2}
\pspolygon[linecolor=white, linewidth=0.02, fillstyle=solid,fillcolor=yellow](0.91751134,1.3317788)(1.9034183,-0.41748416)(2.8348446,1.332738)
\psbezier[linecolor=black, linewidth=0.04, fillstyle=solid,fillcolor=colour0](1.553067,0.89983374)(1.5447986,1.547851)(1.4361526,1.1536276)(1.8759718,1.1576269552442744)(2.315791,1.1616263)(2.0934794,1.5804809)(2.1575115,0.90066475)(2.2215433,0.22084858)(2.306556,0.64546037)(1.8751786,0.54429364)(1.4438012,0.4431269)(1.5613352,0.25181648)(1.553067,0.89983374)
\rput[bl](2.2428446,1.0282936){$M$}
\pscircle[linecolor=black, linewidth=0.02, fillstyle=solid,fillcolor=colour0, dimen=outer](1.7152891,0.4931825){0.05111111}
\pscircle[linecolor=black, linewidth=0.02, fillstyle=solid,fillcolor=colour0, dimen=outer](1.7819558,0.39096028){0.05111111}
\pscircle[linecolor=black, linewidth=0.02, fillstyle=solid,fillcolor=colour0, dimen=outer](1.8930669,0.43984917){0.05111111}
\pscircle[linecolor=black, linewidth=0.02, fillstyle=solid,fillcolor=colour0, dimen=outer](1.884178,0.27540472){0.05111111}
\pscircle[linecolor=black, linewidth=0.02, fillstyle=solid,fillcolor=colour0, dimen=outer](2.0308447,0.40873808){0.05111111}
\pspolygon[linecolor=white, linewidth=0.02, fillstyle=solid,fillcolor=yellow](2.448178,-1.3365248)(1.8933822,-0.34726194)(1.3664002,-1.3508174)
\psline[linecolor=black, linewidth=0.04](0.904178,1.3406092)(1.892704,-0.36415082)(2.8375113,1.3514048)
\pspolygon[linecolor=yellow, linewidth=0.02, fillstyle=solid,fillcolor=yellow](1.953067,-0.35029832)(4.197794,0.09984918)(4.224178,-0.4068175)
\pspolygon[linecolor=yellow, linewidth=0.02, fillstyle=solid,fillcolor=yellow](0.010844693,-0.3490397)(1.8375113,-0.37570637)(0.041955803,-0.7357064)
\psline[linecolor=black, linewidth=0.02](0.0019558037,-0.33570638)(4.2286224,-0.41570637)
\psline[linecolor=black, linewidth=0.02](0.02862247,-0.7490397)(4.197511,0.11318251)
\psline[linecolor=black, linewidth=0.04](2.4064002,-1.331133)(1.8934299,-0.37748414)(1.3575114,-1.3508174)
\pscircle[linecolor=black, linewidth=0.02, fillstyle=solid,fillcolor=colour0, dimen=outer](3.6264002,-0.17348416){0.05111111}
\pscircle[linecolor=black, linewidth=0.02, fillstyle=solid,fillcolor=colour0, dimen=outer](3.7908447,-0.24015082){0.05111111}
\end{pspicture}
}
\caption{A causal fermion system.}
\label{cfs}
\end{figure}

Now one can extend the setting simply by considering~\eqref{Scont} for more general measures
on~$\F$ (like for example regular Borel measures).
The main advantage of working with measures is that we get into a mathematical setting
in which variational principles like~\eqref{Ssum} can be studied with powerful analytic methods.
In particular, as we shall see in Section~\ref{secdiscrete}, in this setting the question
whether minimizers give rise to discrete or continuous space-time structures can be analyzed
and answered.
\eitem
These generalizations lead us to the following definition:
\begin{Def} \label{defcfs} (causal fermion system) {\em{ 
Given a separable complex Hilbert space~$\H$ with scalar product~$\la .|. \ra_\H$
and a parameter~$n \in \N$ (the {\em{``spin dimension''}}), we let~$\F \subset \Lin(\H)$ be the set of all
selfadjoint operators on~$\H$ of finite rank, which (counting multiplicities) have
at most~$n$ positive and at most~$n$ negative eigenvalues. On~$\F$ we are given
a positive measure~$\rho$ (defined on a $\sigma$-algebra of subsets of~$\F$), the so-called
{\em{universal measure}}. We refer to~$(\H, \F, \rho)$ as a {\em{causal fermion system}}.
}}
\end{Def}
This definition is illustrated in Figure~\ref{cfs}.
Now the set~$\F$ is invariant in addition under the transformation where an operator
is multiplied by a negative number, as is indicated in the figure by the double cones.
The support of the measure, denoted by
\beq \label{Mdef}
M := \supp \rho \:,
\eeq
is referred to as {\em{space-time}}. In generalization of the example of the lattice system, where
space-time consisted of discrete points~\eqref{FL}, now the measure~$\rho$ can also have
continuous components.

The Hilbert space~$\H$ can be understood as being spanned by all
wave functions of the system, similar as explained for our lattice system after~\eqref{ortho}.
Indeed, starting from the abstract definition, to every vector~$u \in \H$ one can associate
a corresponding wave function in space-time, the so-called {\em{physical wave function}}
(for details see~\cite[\S1.1.4]{cfs}). In the applications, the physical wave functions 
describe fermions. This is the reason for the name ``causal {\em{fermion}} system.''
The dimension of~$\H$ can be regarded as the number of fermionic particles in the system
(where in the physical applications
we also count the particles of the so-called Dirac sea; see~\cite{srev} for further explanations of this point).

The above notions evolved over several years. Based on preparations in~\cite{pfp},
the present formulation was first given in~\cite{rrev}. For the general background we refer to the
non-technical introduction for physicists in~\cite{dice2014} or to the textbooks~\cite{cfs, intro}.

\section{The Causal Action Principle}
Having explained the general definition of a causal fermion system (see Definition~\ref{defcfs}),
we can now introduce the variational principle used to describe the dynamics of a causal fermion system,
the so-called {\em{causal action principle}}.
The mathematical structure of the causal action is similar to the action~\eqref{Ssum}
given in our example of the lattice system. Its detailed form, however, is far from obvious
and is the result of many computations and long considerations.
The causal action was first proposed in~\cite[Section~3.5]{pfp}.

For any~$x, y \in \F$, the product~$x y$ is an operator of rank at most~$2n$. 
However, in general it is no longer a selfadjoint operator because~$(xy)^* = yx$,
and this is different from~$xy$ unless~$x$ and~$y$ commute.
As a consequence, the eigenvalues of the operator~$xy$ are in general complex.
We denote these eigenvalues counting algebraic multiplicities
by~$\lambda^{xy}_1, \ldots, \lambda^{xy}_{2n} \in \C$
(more specifically,
denoting the rank of~$xy$ by~$k \leq 2n$, we choose~$\lambda^{xy}_1, \ldots, \lambda^{xy}_{k}$ as all
the non-zero eigenvalues and set~$\lambda^{xy}_{k+1}, \ldots, \lambda^{xy}_{2n}=0$).
We introduce the Lagrangian and the causal action by
\begin{align}
\text{\em{Lagrangian:}} && \L(x,y) &= \frac{1}{4n} \sum_{i,j=1}^{2n} \Big( \big|\lambda^{xy}_i \big|
- \big|\lambda^{xy}_j \big| \Big)^2 \label{Lagrange} \\
\text{\em{causal action:}} && \Sact(\rho) &= \iint_{\F \times \F} \L(x,y)\: d\rho(x)\, d\rho(y) \:. \label{Sdef}
\end{align}
The {\em{causal action principle}} is to minimize~$\Sact$ by varying the measure~$\rho$
under the following constraints:
\begin{align}
\text{\em{volume constraint:}} && \rho(\F) = \text{const} \quad\;\; & \label{volconstraint} \\
\text{\em{trace constraint:}} && \int_\F \tr(x)\: d\rho(x) = \text{const}& \label{trconstraint} \\
\text{\em{boundedness constraint:}} && \iint_{\F \times \F} 
\bigg( \sum_{i=1}^{2n} \big|\lambda^{xy}_j \big|^2 \bigg)
\: d\rho(x)\, d\rho(y) &\leq C \:, \label{Tdef}
\end{align}
where~$C$ is a given parameter (and~$\tr$ denotes the trace of a linear operator on~$\H$).
This variational principle is mathematically well-posed if~$\H$ is finite-dimensional and if
one varies the measure in the class of regular Borel measures
(with respect to the topology on~$\Lin(\H)$ induced by the operator norm).
For the existence theory and the analysis of general properties of minimizing measures
we refer to~\cite{discrete, continuum, lagrange}.

Given a minimizing measure~$\rho$, we define {\em{space-time}} again as the support
of the measure~\eqref{Mdef}. Thus the space-time points are selfadjoint linear operators on~$\H$.

The fact that the eigenvalues~$\lambda^{xy}_1, \ldots, \lambda^{xy}_{2n}$ are complex
makes it possible to introduce the following notion of causality:
\begin{Def} (causal structure) \label{def2}
{\em{ The points~$x$ and~$y$ are
called {\em{spacelike}} separated if all the eigenvalues~$\lambda^{xy}_j$ have the same absolute value.
They are said to be {\em{timelike}} separated if the~$\lambda^{xy}_j$ are all real and do not all 
have the same absolute value.
In all other cases (i.e.\ if the~$\lambda^{xy}_j$ are not all real and do not all 
have the same absolute value),
the points~$x$ and~$y$ are said to be {\em{lightlike}} separated. }}
\end{Def} \noindent
Restricting the causal structure of~$\F$ to~$M$, we get causal relations in space-time.

The Lagrangian~\eqref{Lagrange} is compatible with the above notion of causality in the
following sense.
Suppose that two points~$x, y \in \F$ are spacelike separated.
Then the eigenvalues~$\lambda^{xy}_i$ all have the same absolute value.
Therefore, the Lagrangian~\eqref{Lagrange} vanishes. 
Thus pairs of points with spacelike
separation do not enter the action. This can be seen in analogy to the usual notion of causality where
points with spacelike separation cannot influence each other.
This analogy is the reason for the notion {\em{causal}} in ``causal fermion system''
and ``causal action principle.''

We point out that the fact that pairs of points with spacelike separation do not enter the action
is much weaker than the usual concept of causation which states that the present can only
influence the future. Indeed, the structure of the causal Lagrangian leaves the possibility that the
future could influence the present or past. In order to analyze whether this really happens, one
must analyze the initial value problem. If for given initial data there is a unique solution
to the future, we can conclude that causation holds. We will come back to this important
point in Section~\ref{sechyp}.

\section{Discreteness Results} \label{secdiscrete}
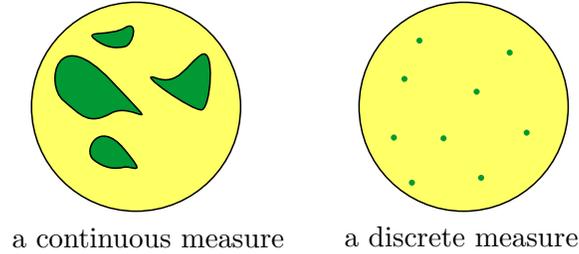
\begin{figure}
%
\psscalebox{1.0 1.0} 
{
\begin{pspicture}(0,-1.635)(7.405,1.635)
\definecolor{colour3}{rgb}{1.0,1.0,0.4}
\definecolor{colour4}{rgb}{0.0,0.6,0.2}
\rput[bl](0.0,-1.635){\normalsize{a continuous measure}}
\pscircle[linecolor=black, linewidth=0.02, fillstyle=solid,fillcolor=colour3, dimen=outer](1.65,0.235){1.4}
\pscircle[linecolor=black, linewidth=0.02, fillstyle=solid,fillcolor=colour3, dimen=outer](6.005,0.235){1.4}
\psbezier[linecolor=black, linewidth=0.02, fillstyle=solid,fillcolor=colour4](1.1,1.135)(0.95,1.255)(1.31,1.17)(1.5,1.285)(1.69,1.4)(1.6163292,1.0611235)(1.5,1.035)(1.3836708,1.0088766)(1.25,1.015)(1.1,1.135)
\psbezier[linecolor=black, linewidth=0.02, fillstyle=solid,fillcolor=colour4](0.68,0.322)(0.44,0.5284)(0.576,1.3712)(1.32,0.58)(2.064,-0.2112)(1.6181266,0.31533232)(1.32,0.15)(1.0218734,-0.01533232)(0.92,0.1156)(0.68,0.322)
\psbezier[linecolor=black, linewidth=0.02, fillstyle=solid,fillcolor=colour4](1.085,-0.47)(0.935,-0.35)(1.155,0.075)(1.485,-0.32)(1.815,-0.715)(1.6013292,-0.5438765)(1.485,-0.57)(1.3686708,-0.59612346)(1.235,-0.59)(1.085,-0.47)
\psbezier[linecolor=black, linewidth=0.02, fillstyle=solid,fillcolor=colour4](1.915,0.6172963)(2.22,0.50785184)(2.321923,0.8006296)(2.514846,0.925)(2.7077692,1.0493704)(2.6431189,0.22325201)(2.525,0.195)(2.406881,0.16674799)(1.61,0.7267407)(1.915,0.6172963)
\pscircle[linecolor=colour4, linewidth=0.02, fillstyle=solid,fillcolor=colour4, dimen=outer](5.4175,1.115){0.04}
\pscircle[linecolor=colour4, linewidth=0.02, fillstyle=solid,fillcolor=colour4, dimen=outer](6.6175,0.955){0.04}
\pscircle[linecolor=colour4, linewidth=0.02, fillstyle=solid,fillcolor=colour4, dimen=outer](5.2175,0.605){0.04}
\pscircle[linecolor=colour4, linewidth=0.02, fillstyle=solid,fillcolor=colour4, dimen=outer](6.2375,-0.71){0.04}
\pscircle[linecolor=colour4, linewidth=0.02, fillstyle=solid,fillcolor=colour4, dimen=outer](6.1775,0.435){0.04}
\pscircle[linecolor=colour4, linewidth=0.02, fillstyle=solid,fillcolor=colour4, dimen=outer](6.8425,-0.11){0.04}
\pscircle[linecolor=colour4, linewidth=0.02, fillstyle=solid,fillcolor=colour4, dimen=outer](5.0775,-0.175){0.04}
\pscircle[linecolor=colour4, linewidth=0.02, fillstyle=solid,fillcolor=colour4, dimen=outer](5.3175,-0.775){0.04}
\pscircle[linecolor=colour4, linewidth=0.02, fillstyle=solid,fillcolor=colour4, dimen=outer](5.7375,-0.185){0.04}
\rput[bl](4.42,-1.62){\normalsize{a discrete measure}}
\end{pspicture}
}
\caption{The causal variational principle on the sphere.}
\label{sphere}
\end{figure}

Clearly, the general setting of causal fermion systems is quite abstract.
In order to explain the causal action principle more concretely,
we now consider the simplest interesting example. 
To this end, we choose the spin dimension~$n=1$
and the Hilbert space~$\H = \C^2$.
Then~$\F$ consists of all Hermitian $2 \times 2$-matrices~$F$ which have at most one positive
and at most one negative eigenvalue. Such a matrix can be represented with the
help of Pauli matrices as
\[ F = \alpha\,1 + \vec{v} \vec{\sigma} \qquad \text{with} \qquad \text{$\alpha \in \R$, $\vec{v}
\in \R^3$ and $|\vec{v}| \geq |\alpha|$} \:. \]
In order to simplify the constraints~\eqref{trconstraint} and~\eqref{Tdef}, we replace them
by the condition that the matrix~$F$ should have eigenvalues~$1\pm \tau$,
where~$\tau \geq 1$ is a parameter of the model
(which plays a similar role as the parameter~$C$ in the boundedness constraint~\eqref{Tdef}).
The set of all matrices with these properties can be written as
\[ \F = \big\{  F = \tau\: \vec{x} \vec{\sigma} + \1 \quad \text{with} \quad \vec{x} \in S^2 \subset \R^3 \big\} \:. \]
Thus the set~$\F$ can be identified with the unit sphere~$S^2$, which also simplifies
illustrations (see Figure~\ref{sphere}).
The volume constraint~\eqref{volconstraint} can be implemented most easily by
restricting attention to {\em{normalized}} regular Borel measures on~$\F$
(i.e.\ measures with~$\rho(\F)=1$). 
Such a measure can be both continuous, discrete
or a mixture. Examples of {\em{continuous measures}} are obtained by
multiplying the Lebesgue measure on the sphere by a non-negative smooth
function. By a {\em{discrete measure}}, on the other hand, we here
mean a {\em{weighted counting measure}},
i.e.\ a measure obtained by inserting weight factors into~\eqref{deltasum},
\beq \label{weighted}
\rho = \sum_{i=1}^L c_i \,\delta_{x_i}
\qquad \text{with} \qquad x_i \in \F\:,\quad c_i \geq 0 \quad \text{and} \quad \sum_{i=1}^L c_i = 1 \:.
\eeq
In this setting, a straightforward computation yields for the Lagrangian~\eqref{Lagrange}
\begin{align*}
\L(x,y) &= \max \big( 0, \D(x,y) \big) 
\qquad \text{with} \\ 
\D(x,y) &= 2 \tau^2\: \big(1+ \langle x,y \rangle \big) \Big( 2 - \tau^2 \: \big(1 - \langle x,y \rangle \big) \Big) \:,
\end{align*}
and~$\langle x,y \rangle$ denotes the Euclidean scalar product of two unit
vectors~$x,y \in S^2 \subset \R^3$
(thus~$\langle x,y \rangle = \cos \vartheta$, where~$\vartheta$ is the
angle between~$x$ and~$y$).
The resulting so-called {\em{causal variational principle on the sphere}}
was introduced in~\cite[Chapter~1]{continuum} and analyzed
in~\cite[Sections~2 and~5]{support} and more recently in~\cite{sphere}.
We now explain a few results from these papers.

First of all, the causal variational principle on the sphere is well-posed,
meaning that the minimum is attained in the class of all normalized regular Borel measures.
The minimizing measure is not unique; indeed, there are typically many minimizers.
The study in~\cite[Section~2]{support} gives the following
\begin{quote}
{\em{numerical result}}: If~$\tau> \sqrt{2}$, every minimizing measure
is a weighted counting measure~\eqref{weighted}.
\end{quote}
Thus, although we minimize over all regular Borel measures
(i.e.\ measures which can have discrete and continuous components),
a minimizing measure always describes a discrete space-time consisting of a finite
number of space-time points (as shown on the right of Figure~\ref{sphere}).
To some extent, this numerical result could be underpinned by analytic
results. First, it was proven in~\cite[Section~5.1]{support} that the
support has an empty interior:
\begin{Thm}
If~$\tau>\sqrt{2}$, the support of any minimizing measure does not
contain an open subset of~$S^2$.
\end{Thm} \noindent
Intuitively speaking, this result shows that the space-time points
are a subset of the sphere of dimension strictly smaller than two.
More recently, it was shown in~\cite{sphere} that the dimension
of the support is even strictly smaller than one:
\begin{Thm}
In the case~$\tau > \sqrt{6}$,
the support of any minimizing measure is totally disconnected and has Hausdorff dimension at most $6/7$.
\end{Thm}
For brevity, we cannot enter the proof of these theorems.
But we point out that the method of proof gives a good understanding
of the underlying mechanism which has the effect that minimizing measures
tend to be supported on low-dimensional subsets of~$\F$.
This mechanism applies not only to the causal variational principle on the sphere, but
similarly also to the general setting in higher dimensions (see~\cite[Section~3.3]{support}).
The gap between the analytic and numerical results comes about mainly due to
shortcomings of the mathematical methods. 

To summarize, the above mathematical results indicate
that minimizing measures of the causal action should be discrete,
and the methods of proof reveal the reason why this should be the case.

\section{Linearized Fields} \label{seclin}
We again point out that the above results give an {\em{indication}} that minimizers of the causal action principle
should correspond to discrete space-times. However, the theorems apply
in too special situations for giving a definitive answer for general causal fermion systems.
Therefore, as mentioned at the beginning of this talk, we
do not want to take sides in favor of discrete or continuous space-times.
Moreover, this distinction does not seem essential to me.
My message is that the fundamental structure
is a measure on~$\F$ which minimizes the causal action.
It is an interesting question whether this measure is discrete or continuous,
but this question seems of secondary importance.

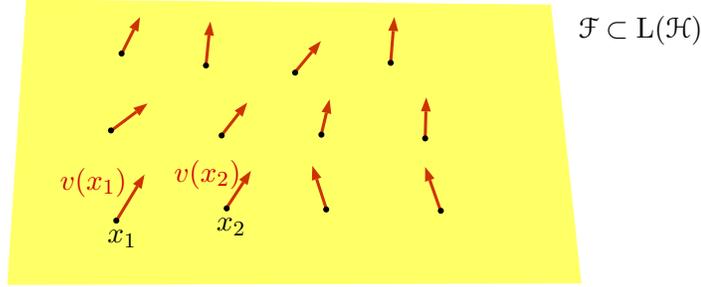
\begin{figure}
%
\psscalebox{1.0 1.0} 
{
\begin{pspicture}(-3,-1.9001338)(11.354513,1.9001338)
\definecolor{colour2}{rgb}{1.0,1.0,0.4}
\definecolor{colour3}{rgb}{0.8,0.2,0.0}
\definecolor{colour4}{rgb}{0.8,0.0,0.0}
\pspolygon[linecolor=colour2, linewidth=0.04, fillstyle=solid,fillcolor=colour2](0.021402044,-1.8800683)(7.611402,-1.8600682)(7.211402,1.8099318)(0.27140203,1.8799318)
\psline[linecolor=colour3, linewidth=0.04, arrowsize=0.05291667cm 2.0,arrowlength=1.4,arrowinset=0.0]{->}(1.526402,1.1849318)(1.766402,1.6649318)
\rput[bl](1.3350687,-1.393846){$x_1$}
\rput[bl](7.594513,1.2939317){$\F \subset \Lin(\H)$}
\psdots[linecolor=black, dotsize=0.08](1.521402,1.1799318)
\psline[linecolor=colour3, linewidth=0.04, arrowsize=0.05291667cm 2.0,arrowlength=1.4,arrowinset=0.0]{->}(1.386402,0.15493175)(1.866402,0.51493174)
\psline[linecolor=colour3, linewidth=0.04, arrowsize=0.05291667cm 2.0,arrowlength=1.4,arrowinset=0.0]{->}(2.861402,0.10493176)(3.191402,0.5249317)
\psline[linecolor=colour3, linewidth=0.04, arrowsize=0.05291667cm 2.0,arrowlength=1.4,arrowinset=0.0]{->}(4.176402,0.10493176)(4.286402,0.56993175)
\psline[linecolor=colour3, linewidth=0.04, arrowsize=0.05291667cm 2.0,arrowlength=1.4,arrowinset=0.0]{->}(5.556402,0.034931757)(5.571402,0.5949318)
\psline[linecolor=colour3, linewidth=0.04, arrowsize=0.05291667cm 2.0,arrowlength=1.4,arrowinset=0.0]{->}(1.4514021,-1.0450683)(1.8264021,-0.43006825)
\psline[linecolor=colour3, linewidth=0.04, arrowsize=0.05291667cm 2.0,arrowlength=1.4,arrowinset=0.0]{->}(2.916402,-0.87506825)(3.2414021,-0.38006824)
\psline[linecolor=colour3, linewidth=0.04, arrowsize=0.05291667cm 2.0,arrowlength=1.4,arrowinset=0.0]{->}(4.241402,-0.88506824)(4.056402,-0.31506824)
\psline[linecolor=colour3, linewidth=0.04, arrowsize=0.05291667cm 2.0,arrowlength=1.4,arrowinset=0.0]{->}(5.7664022,-0.92506826)(5.556402,-0.33006823)
\psline[linecolor=colour3, linewidth=0.04, arrowsize=0.05291667cm 2.0,arrowlength=1.4,arrowinset=0.0]{->}(2.641402,1.0099318)(2.701402,1.6049317)
\psline[linecolor=colour3, linewidth=0.04, arrowsize=0.05291667cm 2.0,arrowlength=1.4,arrowinset=0.0]{->}(3.821402,0.92493176)(4.171402,1.3449317)
\psline[linecolor=colour3, linewidth=0.04, arrowsize=0.05291667cm 2.0,arrowlength=1.4,arrowinset=0.0]{->}(5.096402,1.0399318)(5.146402,1.6649318)
\psdots[linecolor=black, dotsize=0.08](2.641402,1.0199318)
\psdots[linecolor=black, dotsize=0.08](3.826402,0.92493176)
\psdots[linecolor=black, dotsize=0.08](5.096402,1.0549318)
\psdots[linecolor=black, dotsize=0.08](1.381402,0.15493175)
\psdots[linecolor=black, dotsize=0.08](2.851402,0.089931756)
\psdots[linecolor=black, dotsize=0.08](4.176402,0.099931754)
\psdots[linecolor=black, dotsize=0.08](5.556402,0.049931753)
\psdots[linecolor=black, dotsize=0.08](1.4514021,-1.0450683)
\psdots[linecolor=black, dotsize=0.08](2.916402,-0.88006824)
\psdots[linecolor=black, dotsize=0.08](4.241402,-0.8950682)
\psdots[linecolor=black, dotsize=0.08](5.7664022,-0.91506827)
\rput[bl](2.7850688,-1.2388461){$x_2$}
\rput[bl](0.7050687,-0.723846){\textcolor{colour4}{$v(x_1)$}}
\rput[bl](2.2200687,-0.613846){\textcolor{colour4}{$v(x_2)$}}
\end{pspicture}
}
\caption{Jets in a discrete space-time.}
\label{jet}
\end{figure}

Nevertheless, in the remainder of my talk, I will restrict attention to discrete measures
for two reasons: First, because this facilitates the comparison to other discrete approaches
at this conference (like causal sets or cellular automata). Second, 
it has the advantage that integrals reduce to sums, and that the presentation simplifies 
because space-time $M$ can be depicted as a discrete set of points~$x_1, x_2, \ldots \in \F$ (see Figure~\ref{jet}).
We now proceed by explaining a few recent structural results
on causal fermion systems, specialized to discrete measures.

Let us assume that~$\rho$ is a minimizing measure which is discrete
in the sense that its support is a discrete point set~$M = \{x_1, x_2, \ldots\} \subset \F$
(the set~$M$ could be finite or countable, which for our presentation makes no difference).
The fact that~$\rho$ is a minimizer of the causal action poses
many conditions on this measure.
In more physical terms, these conditions tell us about the interaction and the dynamics
of the system. It is an important and difficult task to analyze what these
conditions actually mean, for example by rewriting them
in a form familiar from classical field theory or quantum theory.
In more general terms, the question is:
\begin{quote}
What is the dynamics of a causal fermion system as described
by the causal action principle?
\end{quote}
In order to get detailed and explicit insight into this question, 
the following strategy turns out to be fruitful:
A class of minimizing measures which are relatively easy to construct
are minimizers which have certain symmetries and
can therefore be regarded as describing the vacuum.
Now one can consider small perturbations of the measure~$\rho$
and analyze their dynamics.
Considering linear perturbations~\cite{jet, linhyp}, this procedure is similar to
analyzing linearized fields in Minkowski space.
Linear perturbations are also a suitable starting point for
analyzing nonlinear fields, for example perturbatively~\cite{perturb}.
The procedure of considering small perturbations of a discrete system
bears some similarity to the procedure in solid states physics, where perturbations
of the positions of the atoms of a solid give rise to lattice oscillations described by
phonons. But in our case, of course, the ``ambient space'' in which the discrete points ``live''
is not the three-dimensional Euclidean space or space-time, but instead it is the set~$\F$ of
linear operators on the Hilbert space~$\H$.

Perturbations of the measure~$\rho$ are described
by so-called {\em{jets}}, which we now introduce.
There are two ways to perturb a weighted counting measure~\eqref{weighted}:
one can perturb the weight factors~$c_i$
or the positions~$x_i$ of the space-time points.
The first class of perturbations is described by a real-valued function~$b$ on the
space-time points. The perturbation of the positions~$x_i$, on the other hand,
is described by a vector~$v$ which is tangential to~$\F$ at the point~$x_i$ (see Figure~\ref{jet}).
The jet~$\v$ is defined as the pair of the real-valued function and the vector field,
\[ 
\v = (b, v) \:. \]

The vector field~$v$ appearing in the jet can also be described
more concretely in terms of wave functions and fields in space-time.
In order to get the connection, we first recall that in the example
of the lattice system, we constructed the points in~$\F$ as
local correlation operators of wave functions in space-time~\eqref{matrix}.
A similar construction works for general causal fermion systems,
with the result that every point~$x \in M$ can be recovered as the
local correlation operator of so-called {\em{physical wave functions}}~$\psi_k$, i.e.
\[ x = F(x) \qquad \text{with} \qquad F(x)^j_k= -\Sl \psi_j(x) | \psi_k(x) \Sr_x \:, \]
where~$\Sl .|.\Sr_x$ is the so-called spin scalar product
(for details see~\cite[\S1.1.4]{cfs}).
A vector field~$v$ at~$x$ is a first variation of~$F(x)$, i.e.\
\[ v(x) = \delta F(x) = -\Sl \delta \psi_j(x) \,|\, \psi_k(x) \Sr_x - \Sl \psi_j(x) \,|\, \delta \psi_k(x) \Sr_x \:. \]
In this way, the vector field~$v$ can be described by linear variations of the physical wave functions.
Now one can consider different types of such variations:
\bitem
\item[(a)] One can vary individual physical wave functions. This gives rise to
the so-called {\em{fermionic jets}}.
\item[(b)] Alternatively, one can vary all physical wave functions collectively.
In physical applications when the physical wave functions satisfy the Dirac equation,
such variations can be described by bosonic fields, like for example
\[ \delta \psi_j(x) = -(s_m \,A_j \gamma^j\, \psi_j)(x)\:, \]
where~$s_m$ is a Dirac Green's operator and~$A$ is the electromagnetic potential.
The corresponding jets are referred to as {\em{bosonic jets}}.
\eitem
More detailed explanations on these constructions can be found in~\cite[Sections~6 and~7]{perturb}.

We remark that the connection to bosonic and fermionic fields is worked out in
detail in the so-called {\em{continuum limit}}, which can be thought of as the
limiting case of a lattice system in four-dimensional Minkowski space
when the lattice spacing tends to zero. For brevity, we cannot enter the
methods and results of the continuum limit analysis, but we refer instead to the
textbook~\cite{cfs} or the short survey in~\cite[Section~6]{dice2014}.

\section{The Jet Dynamics} \label{secjet}
The dynamics of linearized fields is described by equations in space-time,
referred to as the {\em{linearized field equations}}. We now outline how these
equations are derived and say a few words on their general structure.
Following the causal action principle, a measure~$\rho$ describing physical space-time
should be a minimizer of the causal action. This implies that~$\rho$ must satisfy
corresponding {\em{Euler-Lagrange equations}}. These equations take the form
(more precisely these are the so-called weak EL equations; for details~\cite[Sections~3.1 or 4.1]{jet})
\beq \label{EL}
\nabla_{\u} \bigg( \int_\F \L(x,y)\: d\rho(y) - \frac{\nu}{2} \bigg) = 0 \:,
\eeq
which must hold for all jets~$\u$ (the so-called {\em{test jets}}) and all~$x \in M$,
where~$\nabla$ denotes the combination of multiplication and differentiation
\[ \nabla_{\v} \eta(x)= b(x) \,\eta (x) + D_v \eta(x) \:. \]
In simple terms, the function in the brackets in~\eqref{EL} as well as its derivatives
must vanish at all space-time points.

We next assume that~$\rho$ describes the vacuum and consider
linear perturbations of the measure~$\rho$ described by a jet~$\v$.
These perturbations should be physical in the sense that the perturbed measures
should also satisfy the Euler-Lagrange equations.
In other words, the jet~$\v$ should generate a one-parameter family of
solutions of~\eqref{EL}. The linearized field equations are
obtained by differentiating the Euler-Lagrange equations with respect to this parameter.
They take the form (for details see~\cite[Section~4.2]{jet})
\[ 
\la \u, \Delta \v \ra(x) = 0 \qquad \text{for all~$x \in M$} \:, \]
where the ``Laplacian'' is defined by
\beq \label{eqlinlip2}
\la \u, \Delta \v \ra(x) := \nabla_{\u} \bigg( \int_M \big( \nabla_{1, \v} + \nabla_{2, \v} \big) \L(x,y)\: d\rho(y) - \nabla_\v \:\frac{\nu}{2} \bigg) \:.
\eeq
For a discrete measure, the last integral reduces to a sum over the space-time points $x_1, x_2, \ldots \in \F$.
Therefore, similar to the wave equation on the lattice~\eqref{wavelattice},
the linearized field equations involve a sum over lattice points.
But, in contrast to~\eqref{wavelattice}, we do not sum only over the nearest neighbors,
but the equation~\eqref{eqlinlip2} is {\em{nonlocal}} in the sense that
it typically involves a sum over many lattice points.

\begin{figure}
%
\psscalebox{1.0 1.0} 
{
\begin{pspicture}(0,-2.2897422)(7.0581255,2.2897422)
\definecolor{colour6}{rgb}{0.6,0.6,1.0}
\definecolor{colour7}{rgb}{0.8,1.0,0.6}
\definecolor{colour5}{rgb}{0.0,0.6,0.2}
\pspolygon[linecolor=colour7, linewidth=0.04, fillstyle=solid,fillcolor=colour7](0.020014497,2.269681)(0.020014497,1.1546808)(0.3950145,0.8446809)(0.9250145,0.28468087)(1.4150145,-0.31031913)(2.1900146,-1.1153191)(2.9250145,-2.1353192)(3.6400144,-2.125319)(4.2000146,-1.4003191)(4.6800146,-1.0003191)(6.0900145,-0.07531912)(6.0900145,-0.07531912)(6.6150146,0.39968088)(6.6000147,2.249681)
\psdots[linecolor=black, dotsize=0.08](0.1600145,1.3996809)
\psdots[linecolor=black, dotsize=0.08](1.3800145,1.3296809)
\psdots[linecolor=black, dotsize=0.08](2.5850146,1.2446809)
\psdots[linecolor=black, dotsize=0.08](3.8950145,1.3046808)
\psdots[linecolor=black, dotsize=0.08](0.21001449,1.1146809)
\rput{98.70961}(4.570297,-1.4441137){\psdots[linecolor=black, dotsize=0.08](2.9050145,1.2396809)}
\psdots[linecolor=black, dotsize=0.08](5.6100144,1.8096809)
\psdots[linecolor=black, dotsize=0.08](3.2100146,0.94968086)
\psdots[linecolor=black, dotsize=0.08](0.6100145,0.7246809)
\psdots[linecolor=black, dotsize=0.08](3.8350146,1.0046809)
\psdots[linecolor=black, dotsize=0.08](3.6100144,1.2496809)
\psdots[linecolor=black, dotsize=0.08](3.5000145,0.9696809)
\psdots[linecolor=black, dotsize=0.08](0.5850145,1.3496809)
\psdots[linecolor=black, dotsize=0.08](1.0350145,1.3146809)
\psdots[linecolor=black, dotsize=0.08](1.6900145,1.2996808)
\psdots[linecolor=black, dotsize=0.08](2.0100145,1.3046808)
\rput{20.632116}(0.5912286,-0.7387842){\psdots[linecolor=black, dotsize=0.08](2.3250146,1.2546809)}
\rput{14.352597}(0.41308272,-0.7714385){\psdots[linecolor=black, dotsize=0.08](3.2700145,1.2546809)}
\psdots[linecolor=black, dotsize=0.08](1.9900146,0.9646809)
\psdots[linecolor=black, dotsize=0.08](2.3200145,0.8796809)
\psdots[linecolor=black, dotsize=0.08](2.6050146,0.8796809)
\psdots[linecolor=black, dotsize=0.08](2.9000144,0.9246809)
\psdots[linecolor=black, dotsize=0.08](0.5300145,1.0396808)
\psdots[linecolor=black, dotsize=0.08](0.8500145,1.0146809)
\psdots[linecolor=black, dotsize=0.08](1.2200145,0.94468087)
\psdots[linecolor=black, dotsize=0.08](1.6200145,0.9796809)
\psdots[linecolor=black, dotsize=0.08](0.2250145,0.8096809)
\psdots[linecolor=black, dotsize=0.08](1.0200145,0.70468086)
\psdots[linecolor=black, dotsize=0.08](1.4800144,0.6046809)
\psdots[linecolor=black, dotsize=0.08](1.9500145,0.5546809)
\psdots[linecolor=black, dotsize=0.08](2.4100144,0.51468086)
\psdots[linecolor=black, dotsize=0.08](2.7900145,0.5546809)
\psdots[linecolor=black, dotsize=0.08](3.1550145,0.5896809)
\psdots[linecolor=black, dotsize=0.08](3.5250144,0.63968086)
\psdots[linecolor=black, dotsize=0.08](3.8750145,0.63968086)
\psdots[linecolor=black, dotsize=0.08](0.1550145,0.40968087)
\psdots[linecolor=black, dotsize=0.08](0.64001447,0.36968088)
\psdots[linecolor=black, dotsize=0.08](1.0750145,0.25468087)
\psdots[linecolor=black, dotsize=0.08](1.5150145,0.21968088)
\psdots[linecolor=black, dotsize=0.08](0.1750145,0.019680882)
\psdots[linecolor=black, dotsize=0.08](0.5750145,-0.04031912)
\psdots[linecolor=black, dotsize=0.08](1.0850145,-0.15031911)
\psdots[linecolor=black, dotsize=0.08](1.5150145,-0.20031911)
\psdots[linecolor=black, dotsize=0.08](1.9450145,-0.28031912)
\psdots[linecolor=black, dotsize=0.08](2.3750145,-0.25031912)
\psdots[linecolor=black, dotsize=0.08](1.9350145,0.17968088)
\psdots[linecolor=black, dotsize=0.08](2.3350146,0.16968088)
\psdots[linecolor=black, dotsize=0.08](2.7550144,0.18968087)
\psdots[linecolor=black, dotsize=0.08](3.1550145,0.23968089)
\psdots[linecolor=black, dotsize=0.08](3.5350144,0.30968088)
\psdots[linecolor=black, dotsize=0.08](3.9550145,0.2696809)
\psdots[linecolor=black, dotsize=0.08](2.7650144,-0.21031912)
\psdots[linecolor=black, dotsize=0.08](3.1450145,-0.18031912)
\psdots[linecolor=black, dotsize=0.08](3.5450144,-0.070319116)
\psdots[linecolor=black, dotsize=0.08](3.9650145,-0.16031912)
\psdots[linecolor=black, dotsize=0.08](0.1450145,-0.49031913)
\psdots[linecolor=black, dotsize=0.08](0.6050145,-0.5603191)
\psdots[linecolor=black, dotsize=0.08](1.1150146,-0.5603191)
\psdots[linecolor=black, dotsize=0.08](1.5550145,-0.6603191)
\psdots[linecolor=black, dotsize=0.08](1.9450145,-0.67031914)
\psdots[linecolor=black, dotsize=0.08](2.4450145,-0.7103191)
\psdots[linecolor=black, dotsize=0.08](2.9550145,-0.6303191)
\psdots[linecolor=black, dotsize=0.08](3.3150146,-0.5603191)
\psdots[linecolor=black, dotsize=0.08](3.6750145,-0.5903191)
\psdots[linecolor=black, dotsize=0.08](4.0750146,-0.60031915)
\psdots[linecolor=black, dotsize=0.08](0.105014496,-0.8903191)
\psdots[linecolor=black, dotsize=0.08](0.5750145,-0.9003191)
\psdots[linecolor=black, dotsize=0.08](1.0350145,-0.9603191)
\psdots[linecolor=black, dotsize=0.08](1.5150145,-1.0203191)
\psdots[linecolor=black, dotsize=0.08](4.1550145,-0.8903191)
\psdots[linecolor=black, dotsize=0.08](3.7250144,-0.8903191)
\psdots[linecolor=black, dotsize=0.08](3.2350144,-0.9403191)
\psdots[linecolor=black, dotsize=0.08](2.7850144,-1.0203191)
\psdots[linecolor=black, dotsize=0.08](2.2950144,-1.0203191)
\psdots[linecolor=black, dotsize=0.08](1.8950145,-1.0003191)
\psdots[linecolor=black, dotsize=0.08](0.1250145,-1.1803191)
\psdots[linecolor=black, dotsize=0.08](0.5650145,-1.2403191)
\psdots[linecolor=black, dotsize=0.08](1.0150145,-1.2703191)
\psdots[linecolor=black, dotsize=0.08](1.5050145,-1.3303192)
\psdots[linecolor=black, dotsize=0.08](2.0350144,-1.3203192)
\psdots[linecolor=black, dotsize=0.08](2.5350144,-1.2903191)
\psdots[linecolor=black, dotsize=0.08](3.0950146,-1.3103191)
\psdots[linecolor=black, dotsize=0.08](3.5150144,-1.2403191)
\psdots[linecolor=black, dotsize=0.08](3.9650145,-1.2203192)
\psdots[linecolor=black, dotsize=0.08](4.2950144,-1.4603192)
\psdots[linecolor=black, dotsize=0.08](3.7550144,-1.5203191)
\psdots[linecolor=black, dotsize=0.08](3.2850144,-1.5603191)
\psdots[linecolor=black, dotsize=0.08](1.2050145,-1.5403191)
\psdots[linecolor=black, dotsize=0.08](0.6450145,-1.5203191)
\psdots[linecolor=black, dotsize=0.08](0.1450145,-1.5103191)
\psdots[linecolor=black, dotsize=0.08](1.7550145,-1.5603191)
\psdots[linecolor=black, dotsize=0.08](2.6850145,-1.5303191)
\psdots[linecolor=black, dotsize=0.08](0.1550145,1.6996809)
\psdots[linecolor=black, dotsize=0.08](0.6250145,1.7296809)
\psdots[linecolor=black, dotsize=0.08](1.0250145,1.6496809)
\psdots[linecolor=black, dotsize=0.08](1.4150145,1.5896809)
\psdots[linecolor=black, dotsize=0.08](1.7850145,1.5496808)
\psdots[linecolor=black, dotsize=0.08](2.1550145,1.5496808)
\psdots[linecolor=black, dotsize=0.08](2.5150144,1.4746809)
\psdots[linecolor=black, dotsize=0.08](2.8950145,1.4446809)
\psdots[linecolor=black, dotsize=0.08](3.1550145,1.6046809)
\psdots[linecolor=black, dotsize=0.08](3.5650146,1.5746809)
\psdots[linecolor=black, dotsize=0.08](3.8850145,1.6046809)
\psdots[linecolor=black, dotsize=0.08](0.4550145,1.9646809)
\psdots[linecolor=black, dotsize=0.08](0.8950145,1.9046808)
\psdots[linecolor=black, dotsize=0.08](1.3350145,1.8446809)
\psdots[linecolor=black, dotsize=0.08](1.7350144,1.8446809)
\psdots[linecolor=black, dotsize=0.08](2.1350145,1.8246809)
\psdots[linecolor=black, dotsize=0.08](2.4950144,1.7046809)
\psdots[linecolor=black, dotsize=0.08](2.8850145,1.8846809)
\psdots[linecolor=black, dotsize=0.08](2.8150146,1.6746808)
\psdots[linecolor=black, dotsize=0.08](2.4350145,1.9896809)
\psdots[linecolor=black, dotsize=0.08](3.2550144,1.9446809)
\psdots[linecolor=black, dotsize=0.08](3.6050146,1.8546809)
\psdots[linecolor=black, dotsize=0.08](3.8950145,1.9546809)
\psdots[linecolor=black, dotsize=0.08](0.075014494,-1.8803191)
\psdots[linecolor=black, dotsize=0.08](0.51501447,-1.8503191)
\psdots[linecolor=black, dotsize=0.08](0.9850145,-1.8003191)
\psdots[linecolor=black, dotsize=0.08](1.4650145,-1.8503191)
\psdots[linecolor=black, dotsize=0.08](1.9550145,-1.8303192)
\psdots[linecolor=black, dotsize=0.08](2.2050145,-1.5703192)
\psdots[linecolor=black, dotsize=0.08](2.4550145,-1.8603191)
\psdots[linecolor=black, dotsize=0.08](2.9050145,-1.7803191)
\psdots[linecolor=black, dotsize=0.08](3.4350145,-1.7903191)
\psdots[linecolor=black, dotsize=0.08](3.8750145,-1.7703191)
\psdots[linecolor=black, dotsize=0.08](4.3350143,-1.7703191)
\psdots[linecolor=black, dotsize=0.08](0.11501449,-2.250319)
\psdots[linecolor=black, dotsize=0.08](0.5650145,-2.200319)
\psdots[linecolor=black, dotsize=0.08](1.0050145,-2.160319)
\psdots[linecolor=black, dotsize=0.08](4.2450147,1.8646809)
\psdots[linecolor=black, dotsize=0.08](4.1550145,1.5946809)
\psdots[linecolor=black, dotsize=0.08](4.1550145,1.2946808)
\psdots[linecolor=black, dotsize=0.08](4.2350144,0.9846809)
\psdots[linecolor=black, dotsize=0.08](4.2650146,0.6646809)
\psdots[linecolor=black, dotsize=0.08](4.3350143,0.24468088)
\psdots[linecolor=black, dotsize=0.08](4.3750143,-0.13531911)
\psdots[linecolor=black, dotsize=0.08](4.4450145,-0.46531913)
\psdots[linecolor=black, dotsize=0.08](4.4850144,-0.8953191)
\psdots[linecolor=black, dotsize=0.08](4.3950143,-1.1553191)
\psdots[linecolor=black, dotsize=0.08](4.6750145,-1.3153191)
\psdots[linecolor=black, dotsize=0.08](4.7250147,-1.6853191)
\psdots[linecolor=black, dotsize=0.08](4.7550144,-2.055319)
\psdots[linecolor=black, dotsize=0.08](4.2950144,-2.065319)
\psdots[linecolor=black, dotsize=0.08](3.9050145,-2.045319)
\psdots[linecolor=black, dotsize=0.08](3.4350145,-2.105319)
\psdots[linecolor=black, dotsize=0.08](3.0150144,-2.065319)
\psdots[linecolor=black, dotsize=0.08](2.5750146,-2.125319)
\psdots[linecolor=black, dotsize=0.08](4.5450144,1.7846808)
\psdots[linecolor=black, dotsize=0.08](4.5050144,1.5346808)
\psdots[linecolor=black, dotsize=0.085](4.5550146,1.2646809)
\psdots[linecolor=black, dotsize=0.08](4.6150146,0.9646809)
\psdots[linecolor=black, dotsize=0.08](4.6250143,0.64468086)
\psdots[linecolor=black, dotsize=0.08](4.7450147,0.2646809)
\psdots[linecolor=black, dotsize=0.08](4.7550144,-0.11531912)
\psdots[linecolor=black, dotsize=0.08](4.8550143,-0.47531912)
\psdots[linecolor=black, dotsize=0.08](4.9400144,-0.7603191)
\psdots[linecolor=black, dotsize=0.08](4.9650145,-1.1053191)
\psdots[linecolor=black, dotsize=0.08](5.0550146,-1.5353191)
\psdots[linecolor=black, dotsize=0.08](5.0650144,-1.8753191)
\psdots[linecolor=black, dotsize=0.08](4.8850145,1.7946808)
\psdots[linecolor=black, dotsize=0.08](4.8350143,1.4746809)
\psdots[linecolor=black, dotsize=0.08](4.9950147,1.0946809)
\psdots[linecolor=black, dotsize=0.08](4.9850144,0.69468087)
\psdots[linecolor=black, dotsize=0.08](5.0750146,0.27468088)
\psdots[linecolor=black, dotsize=0.08](5.1550145,-0.14531912)
\psdots[linecolor=black, dotsize=0.08](5.2450147,-0.5353191)
\psdots[linecolor=black, dotsize=0.08](5.2650146,-0.92531914)
\psdots[linecolor=black, dotsize=0.08](5.2650146,1.7546809)
\psdots[linecolor=black, dotsize=0.08](5.1550145,1.4546809)
\psdots[linecolor=black, dotsize=0.08](5.3350143,1.1146809)
\psdots[linecolor=black, dotsize=0.08](5.3450146,0.7846809)
\psdots[linecolor=black, dotsize=0.08](5.3550143,0.47468087)
\psdots[linecolor=black, dotsize=0.08](5.4150143,0.11468088)
\psdots[linecolor=black, dotsize=0.08](5.4850144,-0.31531912)
\psdots[linecolor=black, dotsize=0.08](5.6050143,-0.7053191)
\psdots[linecolor=black, dotsize=0.08](5.3550143,-1.2153192)
\psdots[linecolor=black, dotsize=0.08](5.6050143,-1.0153191)
\psdots[linecolor=black, dotsize=0.08](5.4050145,-1.5153191)
\psdots[linecolor=black, dotsize=0.08](5.4150143,-1.8353192)
\psdots[linecolor=black, dotsize=0.08](5.6950145,-1.3053191)
\psdots[linecolor=black, dotsize=0.08](5.7650146,-1.6453191)
\psdots[linecolor=black, dotsize=0.08](5.7250147,-2.025319)
\psdots[linecolor=black, dotsize=0.08](5.3250146,-2.105319)
\psdots[linecolor=black, dotsize=0.08](1.7050145,-2.105319)
\psdots[linecolor=black, dotsize=0.08](2.1650145,-2.095319)
\psdots[linecolor=black, dotsize=0.08](5.5350146,1.5046809)
\psdots[linecolor=black, dotsize=0.08](5.6550145,1.1446809)
\psdots[linecolor=black, dotsize=0.08](5.6550145,0.8046809)
\psdots[linecolor=black, dotsize=0.08](5.7150145,0.51468086)
\psdots[linecolor=black, dotsize=0.08](5.7650146,0.17468089)
\psdots[linecolor=black, dotsize=0.08](5.8150144,-0.16531911)
\psdots[linecolor=black, dotsize=0.08](5.8550143,-0.4853191)
\psdots[linecolor=black, dotsize=0.08](5.9950147,-0.80531913)
\psdots[linecolor=black, dotsize=0.08](6.0050144,-1.0553191)
\psdots[linecolor=black, dotsize=0.08](6.0650144,-1.4053191)
\psdots[linecolor=black, dotsize=0.08](6.1050143,-1.7453191)
\psdots[linecolor=black, dotsize=0.08](6.1450143,-2.015319)
\psdots[linecolor=black, dotsize=0.08](5.9350147,1.7846808)
\psdots[linecolor=black, dotsize=0.08](5.9050145,1.5346808)
\psdots[linecolor=black, dotsize=0.08](6.0350146,1.1846809)
\psdots[linecolor=black, dotsize=0.08](6.0350146,0.8646809)
\psdots[linecolor=black, dotsize=0.08](6.0650144,0.47468087)
\psdots[linecolor=black, dotsize=0.0875](6.1550145,0.124680884)
\psdots[linecolor=black, dotsize=0.08](6.1950145,-0.19531912)
\psdots[linecolor=black, dotsize=0.08](6.3350143,1.7546809)
\psdots[linecolor=black, dotsize=0.08](6.2650146,1.4646809)
\psdots[linecolor=black, dotsize=0.08](6.3550143,1.0646809)
\psdots[linecolor=black, dotsize=0.08](6.4050145,0.7146809)
\psdots[linecolor=black, dotsize=0.08](6.4450145,0.3946809)
\psdots[linecolor=black, dotsize=0.08](6.4950147,0.05468088)
\psdots[linecolor=black, dotsize=0.08](6.5650144,-0.28531912)
\psdots[linecolor=black, dotsize=0.08](6.2650146,-0.5353191)
\psdots[linecolor=black, dotsize=0.08](6.7150145,-0.6453191)
\psdots[linecolor=black, dotsize=0.08](6.3950143,-0.80531913)
\psdots[linecolor=black, dotsize=0.08](6.4250145,-1.1353191)
\psdots[linecolor=black, dotsize=0.08](6.4450145,-1.4353191)
\psdots[linecolor=black, dotsize=0.08](6.5050144,-1.7353191)
\psdots[linecolor=black, dotsize=0.08](6.6250143,-2.085319)
\psdots[linecolor=black, dotsize=0.08](6.7450147,-1.0353191)
\psdots[linecolor=black, dotsize=0.085](6.8550143,-1.3853191)
\psdots[linecolor=black, dotsize=0.08](6.9250145,-1.6453191)
\psdots[linecolor=black, dotsize=0.08](6.9450145,-1.9353191)
\psdots[linecolor=black, dotsize=0.08](0.075014494,1.9746809)
\psline[linecolor=colour5, linewidth=0.04, arrowsize=0.05291667cm 2.0,arrowlength=1.4,arrowinset=0.0]{->}(3.0150144,-2.0703192)(3.1100144,-1.7353191)
\psline[linecolor=colour5, linewidth=0.04, arrowsize=0.05291667cm 2.0,arrowlength=1.4,arrowinset=0.0]{->}(3.4300146,-2.105319)(3.5850146,-1.8053191)
\psline[linecolor=colour5, linewidth=0.04, arrowsize=0.05291667cm 2.0,arrowlength=1.4,arrowinset=0.0]{->}(2.9200144,-1.7753191)(2.9250145,-1.3453192)
\psline[linecolor=colour5, linewidth=0.04, arrowsize=0.05291667cm 2.0,arrowlength=1.4,arrowinset=0.0]{->}(3.9650145,-0.16031912)(4.1850147,0.2696809)
\psline[linecolor=colour5, linewidth=0.04, arrowsize=0.05291667cm 2.0,arrowlength=1.4,arrowinset=0.0]{->}(3.4400146,-1.7803191)(3.6600144,-1.3153191)
\psline[linecolor=colour5, linewidth=0.04, arrowsize=0.05291667cm 2.0,arrowlength=1.4,arrowinset=0.0]{->}(3.8700144,-1.7803191)(4.1000147,-1.3853191)
\psline[linecolor=colour5, linewidth=0.04, arrowsize=0.05291667cm 2.0,arrowlength=1.4,arrowinset=0.0]{->}(3.5500145,-0.070319116)(3.7000146,0.3596809)
\psline[linecolor=colour5, linewidth=0.04, arrowsize=0.05291667cm 2.0,arrowlength=1.4,arrowinset=0.0]{->}(4.4350147,-0.47531912)(4.6300144,-0.025319118)
\psline[linecolor=colour5, linewidth=0.04, arrowsize=0.05291667cm 2.0,arrowlength=1.4,arrowinset=0.0]{->}(2.6800146,-1.5353191)(2.7350144,-1.2103192)
\psline[linecolor=colour5, linewidth=0.04, arrowsize=0.05291667cm 2.0,arrowlength=1.4,arrowinset=0.0]{->}(3.2850144,-1.5603191)(3.3650146,-1.2153192)
\psline[linecolor=colour5, linewidth=0.04, arrowsize=0.05291667cm 2.0,arrowlength=1.4,arrowinset=0.0]{->}(3.0900145,-1.3203192)(3.1250145,-0.9403191)
\psline[linecolor=colour5, linewidth=0.04, arrowsize=0.05291667cm 2.0,arrowlength=1.4,arrowinset=0.0]{->}(3.5250144,-1.2253191)(3.8300145,-0.9553191)
\psline[linecolor=colour5, linewidth=0.04, arrowsize=0.05291667cm 2.0,arrowlength=1.4,arrowinset=0.0]{->}(3.7550144,-1.5153191)(3.9250145,-1.2153192)
\psline[linecolor=colour5, linewidth=0.04, arrowsize=0.05291667cm 2.0,arrowlength=1.4,arrowinset=0.0]{->}(2.5350144,-1.2903191)(2.4750144,-0.9553191)
\psline[linecolor=colour5, linewidth=0.04, arrowsize=0.05291667cm 2.0,arrowlength=1.4,arrowinset=0.0]{->}(2.7900145,-1.0153191)(2.7000146,-0.61031914)
\psline[linecolor=colour5, linewidth=0.04, arrowsize=0.05291667cm 2.0,arrowlength=1.4,arrowinset=0.0]{->}(3.2350144,-0.9403191)(3.4050145,-0.6403191)
\psline[linecolor=colour5, linewidth=0.04, arrowsize=0.05291667cm 2.0,arrowlength=1.4,arrowinset=0.0]{->}(2.3100145,-1.0103191)(2.1900146,-0.67531914)
\psline[linecolor=colour5, linewidth=0.04, arrowsize=0.05291667cm 2.0,arrowlength=1.4,arrowinset=0.0]{->}(1.9400145,-0.28031912)(1.8750145,0.06468088)
\psline[linecolor=colour5, linewidth=0.04, arrowsize=0.05291667cm 2.0,arrowlength=1.4,arrowinset=0.0]{->}(3.1450145,-0.19031912)(3.2850144,0.17468089)
\psline[linecolor=colour5, linewidth=0.04, arrowsize=0.05291667cm 2.0,arrowlength=1.4,arrowinset=0.0]{->}(3.9500146,-1.2303191)(4.1200147,-0.93031913)
\psline[linecolor=colour5, linewidth=0.04, arrowsize=0.05291667cm 2.0,arrowlength=1.4,arrowinset=0.0]{->}(3.7150145,-0.9053191)(3.9350145,-0.5853191)
\psline[linecolor=colour5, linewidth=0.04, arrowsize=0.05291667cm 2.0,arrowlength=1.4,arrowinset=0.0]{->}(4.3950143,-1.1503191)(4.7250147,-0.8203191)
\psline[linecolor=colour5, linewidth=0.04, arrowsize=0.05291667cm 2.0,arrowlength=1.4,arrowinset=0.0]{->}(4.4800143,-0.9003191)(4.7450147,-0.5903191)
\psline[linecolor=colour5, linewidth=0.04, arrowsize=0.05291667cm 2.0,arrowlength=1.4,arrowinset=0.0]{->}(4.1450143,-0.9103191)(4.2950144,-0.5103191)
\psline[linecolor=colour5, linewidth=0.04, arrowsize=0.05291667cm 2.0,arrowlength=1.4,arrowinset=0.0]{->}(4.0650144,-0.61031914)(4.2600145,-0.12531912)
\psline[linecolor=colour5, linewidth=0.04, arrowsize=0.05291667cm 2.0,arrowlength=1.4,arrowinset=0.0]{->}(3.6750145,-0.5903191)(3.8750145,-0.19531912)
\psline[linecolor=colour5, linewidth=0.04, arrowsize=0.05291667cm 2.0,arrowlength=1.4,arrowinset=0.0]{->}(0.1550145,1.6896809)(0.105014496,1.9196808)
\psline[linecolor=colour5, linewidth=0.04, arrowsize=0.05291667cm 2.0,arrowlength=1.4,arrowinset=0.0]{->}(0.5850145,1.3396809)(0.5700145,1.6646808)
\psline[linecolor=colour5, linewidth=0.04, arrowsize=0.05291667cm 2.0,arrowlength=1.4,arrowinset=0.0]{->}(0.8500145,1.0096809)(0.7700145,1.3046808)
\psline[linecolor=colour5, linewidth=0.04, arrowsize=0.05291667cm 2.0,arrowlength=1.4,arrowinset=0.0]{->}(1.0100145,0.6896809)(1.0500145,1.0596809)
\psline[linecolor=colour5, linewidth=0.04, arrowsize=0.05291667cm 2.0,arrowlength=1.4,arrowinset=0.0]{->}(1.0750145,0.24968088)(0.9500145,0.6596809)
\psline[linecolor=colour5, linewidth=0.04, arrowsize=0.05291667cm 2.0,arrowlength=1.4,arrowinset=0.0]{->}(1.5150145,-0.20531912)(1.4550145,0.15468088)
\psline[linecolor=colour5, linewidth=0.04, arrowsize=0.05291667cm 2.0,arrowlength=1.4,arrowinset=0.0]{->}(2.3750145,-0.25531912)(2.2250144,0.15468088)
\psline[linecolor=colour5, linewidth=0.04, arrowsize=0.05291667cm 2.0,arrowlength=1.4,arrowinset=0.0]{->}(1.9400145,-0.67531914)(1.8200145,-0.28031912)
\psline[linecolor=colour5, linewidth=0.04, arrowsize=0.05291667cm 2.0,arrowlength=1.4,arrowinset=0.0]{->}(3.3100145,-0.5603191)(3.4800146,-0.2603191)
\psline[linecolor=colour5, linewidth=0.04, arrowsize=0.05291667cm 2.0,arrowlength=1.4,arrowinset=0.0]{->}(2.9500146,-0.6303191)(3.0050144,-0.22031912)
\psline[linecolor=colour5, linewidth=0.04, arrowsize=0.05291667cm 2.0,arrowlength=1.4,arrowinset=0.0]{->}(2.4500146,-0.7103191)(2.4950144,-0.3003191)
\psline[linecolor=colour5, linewidth=0.04, arrowsize=0.05291667cm 2.0,arrowlength=1.4,arrowinset=0.0]{->}(3.8950145,1.9446809)(4.0650144,2.244681)
\psline[linecolor=colour5, linewidth=0.04, arrowsize=0.05291667cm 2.0,arrowlength=1.4,arrowinset=0.0]{->}(0.6100145,0.7146809)(0.4050145,1.0246809)
\psline[linecolor=colour5, linewidth=0.04, arrowsize=0.05291667cm 2.0,arrowlength=1.4,arrowinset=0.0]{->}(2.7650144,-0.21031912)(2.8700144,0.13968088)
\psline[linecolor=colour5, linewidth=0.04, arrowsize=0.05291667cm 2.0,arrowlength=1.4,arrowinset=0.0]{->}(2.3350146,0.17468089)(2.2550144,0.50968087)
\psline[linecolor=colour5, linewidth=0.04, arrowsize=0.05291667cm 2.0,arrowlength=1.4,arrowinset=0.0]{->}(1.9350145,0.18468088)(1.7650145,0.5396809)
\psline[linecolor=colour5, linewidth=0.04, arrowsize=0.05291667cm 2.0,arrowlength=1.4,arrowinset=0.0]{->}(1.5200145,0.21968088)(1.4100145,0.5296809)
\psline[linecolor=colour5, linewidth=0.04, arrowsize=0.05291667cm 2.0,arrowlength=1.4,arrowinset=0.0]{->}(0.6200145,1.7246809)(0.6950145,1.9846809)
\psline[linecolor=colour5, linewidth=0.04, arrowsize=0.05291667cm 2.0,arrowlength=1.4,arrowinset=0.0]{->}(0.44001448,1.9346809)(0.4350145,2.244681)
\psline[linecolor=colour5, linewidth=0.04, arrowsize=0.05291667cm 2.0,arrowlength=1.4,arrowinset=0.0]{->}(0.075014494,1.9646809)(0.045014497,2.2646809)
\psline[linecolor=colour5, linewidth=0.04, arrowsize=0.05291667cm 2.0,arrowlength=1.4,arrowinset=0.0]{->}(0.1550145,1.3946809)(0.030014496,1.6996809)
\psline[linecolor=colour5, linewidth=0.04, arrowsize=0.05291667cm 2.0,arrowlength=1.4,arrowinset=0.0]{->}(2.7500145,0.17968088)(2.6950145,0.51468086)
\psline[linecolor=colour5, linewidth=0.04, arrowsize=0.05291667cm 2.0,arrowlength=1.4,arrowinset=0.0]{->}(1.4150145,1.5896809)(1.5850145,1.8896809)
\psline[linecolor=colour5, linewidth=0.04, arrowsize=0.05291667cm 2.0,arrowlength=1.4,arrowinset=0.0]{->}(1.3300145,1.8446809)(1.4400145,2.2096808)
\psline[linecolor=colour5, linewidth=0.04, arrowsize=0.05291667cm 2.0,arrowlength=1.4,arrowinset=0.0]{->}(1.0450145,1.3146809)(1.1000144,1.5996809)
\psline[linecolor=colour5, linewidth=0.04, arrowsize=0.05291667cm 2.0,arrowlength=1.4,arrowinset=0.0]{->}(1.0300145,1.6496809)(1.1250145,1.9896809)
\psline[linecolor=colour5, linewidth=0.04, arrowsize=0.05291667cm 2.0,arrowlength=1.4,arrowinset=0.0]{->}(0.8950145,1.9046808)(1.0100145,2.214681)
\psline[linecolor=colour5, linewidth=0.04, arrowsize=0.05291667cm 2.0,arrowlength=1.4,arrowinset=0.0]{->}(1.6150146,0.9846809)(1.5350145,1.3246809)
\psline[linecolor=colour5, linewidth=0.04, arrowsize=0.05291667cm 2.0,arrowlength=1.4,arrowinset=0.0]{->}(1.4800144,0.6096809)(1.4350145,0.9696809)
\psline[linecolor=colour5, linewidth=0.04, arrowsize=0.05291667cm 2.0,arrowlength=1.4,arrowinset=0.0]{->}(1.2250144,0.94968086)(1.2750145,1.3196809)
\psline[linecolor=colour5, linewidth=0.04, arrowsize=0.05291667cm 2.0,arrowlength=1.4,arrowinset=0.0]{->}(1.3800145,1.3296809)(1.3000145,1.6696808)
\psline[linecolor=colour5, linewidth=0.04, arrowsize=0.05291667cm 2.0,arrowlength=1.4,arrowinset=0.0]{->}(3.1500144,0.58468086)(3.3200145,0.88468087)
\psline[linecolor=colour5, linewidth=0.04, arrowsize=0.05291667cm 2.0,arrowlength=1.4,arrowinset=0.0]{->}(3.1500144,0.23468088)(3.2150145,0.5396809)
\psline[linecolor=colour5, linewidth=0.04, arrowsize=0.05291667cm 2.0,arrowlength=1.4,arrowinset=0.0]{->}(2.7750144,0.5396809)(2.8350146,0.83468086)
\psline[linecolor=colour5, linewidth=0.04, arrowsize=0.05291667cm 2.0,arrowlength=1.4,arrowinset=0.0]{->}(2.4050145,0.51968086)(2.4600146,0.82468086)
\psline[linecolor=colour5, linewidth=0.04, arrowsize=0.05291667cm 2.0,arrowlength=1.4,arrowinset=0.0]{->}(1.9450145,0.5496809)(1.9100145,0.88468087)
\psline[linecolor=colour5, linewidth=0.04, arrowsize=0.05291667cm 2.0,arrowlength=1.4,arrowinset=0.0]{->}(3.2000146,0.94968086)(3.1750145,1.3146809)
\psline[linecolor=colour5, linewidth=0.04, arrowsize=0.05291667cm 2.0,arrowlength=1.4,arrowinset=0.0]{->}(2.8950145,0.9196809)(2.9800146,1.1946809)
\psline[linecolor=colour5, linewidth=0.04, arrowsize=0.05291667cm 2.0,arrowlength=1.4,arrowinset=0.0]{->}(2.6050146,0.8796809)(2.7000146,1.2146809)
\psline[linecolor=colour5, linewidth=0.04, arrowsize=0.05291667cm 2.0,arrowlength=1.4,arrowinset=0.0]{->}(2.3100145,0.8696809)(2.3600144,1.1896809)
\psline[linecolor=colour5, linewidth=0.04, arrowsize=0.05291667cm 2.0,arrowlength=1.4,arrowinset=0.0]{->}(1.9900146,0.9746809)(1.9150145,1.3046808)
\psline[linecolor=colour5, linewidth=0.04, arrowsize=0.05291667cm 2.0,arrowlength=1.4,arrowinset=0.0]{->}(1.6900145,1.2946808)(1.7150145,1.6446809)
\psline[linecolor=colour5, linewidth=0.04, arrowsize=0.05291667cm 2.0,arrowlength=1.4,arrowinset=0.0]{->}(2.1400144,1.5346808)(2.3100145,1.8346809)
\psline[linecolor=colour5, linewidth=0.04, arrowsize=0.05291667cm 2.0,arrowlength=1.4,arrowinset=0.0]{->}(1.9900146,1.2896808)(2.0200145,1.6596808)
\psline[linecolor=colour5, linewidth=0.04, arrowsize=0.05291667cm 2.0,arrowlength=1.4,arrowinset=0.0]{->}(2.4250145,1.9896809)(2.5950146,2.289681)
\psline[linecolor=colour5, linewidth=0.04, arrowsize=0.05291667cm 2.0,arrowlength=1.4,arrowinset=0.0]{->}(2.1250145,1.8096809)(2.2100146,2.1896808)
\psline[linecolor=colour5, linewidth=0.04, arrowsize=0.05291667cm 2.0,arrowlength=1.4,arrowinset=0.0]{->}(1.7350144,1.8496809)(1.7550145,2.1896808)
\psline[linecolor=colour5, linewidth=0.04, arrowsize=0.05291667cm 2.0,arrowlength=1.4,arrowinset=0.0]{->}(1.7700145,1.5396808)(1.9400145,1.8396809)
\psline[linecolor=colour5, linewidth=0.04, arrowsize=0.05291667cm 2.0,arrowlength=1.4,arrowinset=0.0]{->}(2.5800145,1.2446809)(2.7500145,1.5446808)
\psline[linecolor=colour5, linewidth=0.04, arrowsize=0.05291667cm 2.0,arrowlength=1.4,arrowinset=0.0]{->}(2.3200145,1.2596809)(2.4900146,1.5596809)
\psline[linecolor=colour5, linewidth=0.04, arrowsize=0.05291667cm 2.0,arrowlength=1.4,arrowinset=0.0]{->}(3.8750145,0.63968086)(3.9900146,0.94468087)
\psline[linecolor=colour5, linewidth=0.04, arrowsize=0.05291667cm 2.0,arrowlength=1.4,arrowinset=0.0]{->}(3.8300145,0.9996809)(3.9450145,1.2696809)
\psline[linecolor=colour5, linewidth=0.04, arrowsize=0.05291667cm 2.0,arrowlength=1.4,arrowinset=0.0]{->}(3.4950144,0.9746809)(3.5200145,1.2946808)
\psline[linecolor=colour5, linewidth=0.04, arrowsize=0.05291667cm 2.0,arrowlength=1.4,arrowinset=0.0]{->}(3.5300145,0.63968086)(3.7000146,0.9396809)
\psline[linecolor=colour5, linewidth=0.04, arrowsize=0.05291667cm 2.0,arrowlength=1.4,arrowinset=0.0]{->}(4.3750143,-0.13531911)(4.5450144,0.16468088)
\psline[linecolor=colour5, linewidth=0.04, arrowsize=0.05291667cm 2.0,arrowlength=1.4,arrowinset=0.0]{->}(4.3300147,0.23968089)(4.5000143,0.5396809)
\psline[linecolor=colour5, linewidth=0.04, arrowsize=0.05291667cm 2.0,arrowlength=1.4,arrowinset=0.0]{->}(3.9400146,0.23468088)(4.0850143,0.6096809)
\psline[linecolor=colour5, linewidth=0.04, arrowsize=0.05291667cm 2.0,arrowlength=1.4,arrowinset=0.0]{->}(3.5350144,0.31468087)(3.6650145,0.6246809)
\psline[linecolor=colour5, linewidth=0.04, arrowsize=0.05291667cm 2.0,arrowlength=1.4,arrowinset=0.0]{->}(3.8850145,1.2896808)(3.9600146,1.5796809)
\psline[linecolor=colour5, linewidth=0.04, arrowsize=0.05291667cm 2.0,arrowlength=1.4,arrowinset=0.0]{->}(3.6150146,1.2496809)(3.6500144,1.5846809)
\psline[linecolor=colour5, linewidth=0.04, arrowsize=0.05291667cm 2.0,arrowlength=1.4,arrowinset=0.0]{->}(3.5650146,1.5746809)(3.7350144,1.8746809)
\psline[linecolor=colour5, linewidth=0.04, arrowsize=0.05291667cm 2.0,arrowlength=1.4,arrowinset=0.0]{->}(3.5950146,1.8446809)(3.7050145,2.2196808)
\psline[linecolor=colour5, linewidth=0.04, arrowsize=0.05291667cm 2.0,arrowlength=1.4,arrowinset=0.0]{->}(3.2500145,1.9396809)(3.3450146,2.2396808)
\psline[linecolor=colour5, linewidth=0.04, arrowsize=0.05291667cm 2.0,arrowlength=1.4,arrowinset=0.0]{->}(3.1500144,1.5996809)(3.1900146,1.9096808)
\psline[linecolor=colour5, linewidth=0.04, arrowsize=0.05291667cm 2.0,arrowlength=1.4,arrowinset=0.0]{->}(3.2600145,1.2446809)(3.3700144,1.6396809)
\psline[linecolor=colour5, linewidth=0.04, arrowsize=0.05291667cm 2.0,arrowlength=1.4,arrowinset=0.0]{->}(2.9050145,1.2496809)(3.0750146,1.5496808)
\psline[linecolor=colour5, linewidth=0.04, arrowsize=0.05291667cm 2.0,arrowlength=1.4,arrowinset=0.0]{->}(2.8900144,1.4396809)(3.0600145,1.7396809)
\psline[linecolor=colour5, linewidth=0.04, arrowsize=0.05291667cm 2.0,arrowlength=1.4,arrowinset=0.0]{->}(2.8750145,1.8746809)(2.9900146,2.2196808)
\psline[linecolor=colour5, linewidth=0.04, arrowsize=0.05291667cm 2.0,arrowlength=1.4,arrowinset=0.0]{->}(2.8150146,1.6696808)(2.9900146,1.9446809)
\psline[linecolor=colour5, linewidth=0.04, arrowsize=0.05291667cm 2.0,arrowlength=1.4,arrowinset=0.0]{->}(2.4900146,1.6996809)(2.6600144,1.9996809)
\psline[linecolor=colour5, linewidth=0.04, arrowsize=0.05291667cm 2.0,arrowlength=1.4,arrowinset=0.0]{->}(3.8900144,1.6196809)(4.0600147,1.9196808)
\psline[linecolor=colour5, linewidth=0.04, arrowsize=0.05291667cm 2.0,arrowlength=1.4,arrowinset=0.0]{->}(5.2650146,1.7546809)(5.3800144,2.079681)
\psline[linecolor=colour5, linewidth=0.04, arrowsize=0.05291667cm 2.0,arrowlength=1.4,arrowinset=0.0]{->}(5.4200144,0.11968088)(5.5900145,0.4196809)
\psline[linecolor=colour5, linewidth=0.04, arrowsize=0.05291667cm 2.0,arrowlength=1.4,arrowinset=0.0]{->}(5.1550145,-0.14531912)(5.3450146,0.25468087)
\psline[linecolor=colour5, linewidth=0.04, arrowsize=0.05291667cm 2.0,arrowlength=1.4,arrowinset=0.0]{->}(4.8550143,-0.47531912)(5.0250144,-0.17531912)
\psline[linecolor=colour5, linewidth=0.04, arrowsize=0.05291667cm 2.0,arrowlength=1.4,arrowinset=0.0]{->}(4.9500146,-0.7503191)(5.1200147,-0.4503191)
\psline[linecolor=colour5, linewidth=0.04, arrowsize=0.05291667cm 2.0,arrowlength=1.4,arrowinset=0.0]{->}(6.1500144,0.11468088)(6.2750144,0.3646809)
\psline[linecolor=colour5, linewidth=0.04, arrowsize=0.05291667cm 2.0,arrowlength=1.4,arrowinset=0.0]{->}(5.8100147,-0.17031913)(5.9900146,0.18968087)
\psline[linecolor=colour5, linewidth=0.04, arrowsize=0.05291667cm 2.0,arrowlength=1.4,arrowinset=0.0]{->}(5.4850144,-0.3003191)(5.6250143,-0.06531912)
\psline[linecolor=colour5, linewidth=0.04, arrowsize=0.05291667cm 2.0,arrowlength=1.4,arrowinset=0.0]{->}(5.2550144,-0.5203191)(5.3950143,-0.2603191)
\psline[linecolor=colour5, linewidth=0.04, arrowsize=0.05291667cm 2.0,arrowlength=1.4,arrowinset=0.0]{->}(5.0700145,0.27468088)(5.2550144,0.70468086)
\psline[linecolor=colour5, linewidth=0.04, arrowsize=0.05291667cm 2.0,arrowlength=1.4,arrowinset=0.0]{->}(4.7400146,0.25968087)(4.9100146,0.5596809)
\psline[linecolor=colour5, linewidth=0.04, arrowsize=0.05291667cm 2.0,arrowlength=1.4,arrowinset=0.0]{->}(4.7500143,-0.12031912)(4.9200144,0.17968088)
\psline[linecolor=colour5, linewidth=0.04, arrowsize=0.05291667cm 2.0,arrowlength=1.4,arrowinset=0.0]{->}(6.3550143,1.0696809)(6.4900146,1.3996809)
\psline[linecolor=colour5, linewidth=0.04, arrowsize=0.05291667cm 2.0,arrowlength=1.4,arrowinset=0.0]{->}(5.7650146,0.18468088)(6.0250144,0.6796809)
\psline[linecolor=colour5, linewidth=0.04, arrowsize=0.05291667cm 2.0,arrowlength=1.4,arrowinset=0.0]{->}(6.0600147,0.46968088)(6.2900143,0.9146809)
\psline[linecolor=colour5, linewidth=0.04, arrowsize=0.05291667cm 2.0,arrowlength=1.4,arrowinset=0.0]{->}(6.4000144,0.7196809)(6.5300145,1.0496808)
\psline[linecolor=colour5, linewidth=0.04, arrowsize=0.05291667cm 2.0,arrowlength=1.4,arrowinset=0.0]{->}(6.4400144,0.3896809)(6.5550146,0.6846809)
\psline[linecolor=colour5, linewidth=0.04, arrowsize=0.05291667cm 2.0,arrowlength=1.4,arrowinset=0.0]{->}(5.3400145,0.7896809)(5.5200143,1.1746808)
\psline[linecolor=colour5, linewidth=0.04, arrowsize=0.05291667cm 2.0,arrowlength=1.4,arrowinset=0.0]{->}(4.9900146,0.69968086)(5.0700145,0.9996809)
\psline[linecolor=colour5, linewidth=0.04, arrowsize=0.05291667cm 2.0,arrowlength=1.4,arrowinset=0.0]{->}(4.6150146,0.63968086)(4.7250147,0.9146809)
\psline[linecolor=colour5, linewidth=0.04, arrowsize=0.05291667cm 2.0,arrowlength=1.4,arrowinset=0.0]{->}(4.2500143,0.6546809)(4.4400144,1.0946809)
\psline[linecolor=colour5, linewidth=0.04, arrowsize=0.05291667cm 2.0,arrowlength=1.4,arrowinset=0.0]{->}(6.3250146,1.7346809)(6.4950147,2.0346808)
\psline[linecolor=colour5, linewidth=0.04, arrowsize=0.05291667cm 2.0,arrowlength=1.4,arrowinset=0.0]{->}(6.2700143,1.4646809)(6.4400144,1.7646809)
\psline[linecolor=colour5, linewidth=0.04, arrowsize=0.05291667cm 2.0,arrowlength=1.4,arrowinset=0.0]{->}(6.0300145,1.1746808)(6.2000146,1.4746809)
\psline[linecolor=colour5, linewidth=0.04, arrowsize=0.05291667cm 2.0,arrowlength=1.4,arrowinset=0.0]{->}(5.6500144,0.7946809)(5.9500146,1.3746809)
\psline[linecolor=colour5, linewidth=0.04, arrowsize=0.05291667cm 2.0,arrowlength=1.4,arrowinset=0.0]{->}(5.3500147,0.46468088)(5.5550146,0.8696809)
\psline[linecolor=colour5, linewidth=0.04, arrowsize=0.05291667cm 2.0,arrowlength=1.4,arrowinset=0.0]{->}(6.0400143,0.8646809)(6.2500143,1.1596808)
\psline[linecolor=colour5, linewidth=0.04, arrowsize=0.05291667cm 2.0,arrowlength=1.4,arrowinset=0.0]{->}(5.7100143,0.5046809)(5.8800144,0.7546809)
\psline[linecolor=colour5, linewidth=0.04, arrowsize=0.05291667cm 2.0,arrowlength=1.4,arrowinset=0.0]{->}(4.8250146,1.4646809)(4.9950147,1.7646809)
\psline[linecolor=colour5, linewidth=0.04, arrowsize=0.05291667cm 2.0,arrowlength=1.4,arrowinset=0.0]{->}(4.5000143,1.5296808)(4.6700144,1.8296809)
\psline[linecolor=colour5, linewidth=0.04, arrowsize=0.05291667cm 2.0,arrowlength=1.4,arrowinset=0.0]{->}(4.1450143,1.5896809)(4.3150144,1.8896809)
\psline[linecolor=colour5, linewidth=0.04, arrowsize=0.05291667cm 2.0,arrowlength=1.4,arrowinset=0.0]{->}(4.5500145,1.2596809)(4.6700144,1.5746809)
\psline[linecolor=colour5, linewidth=0.04, arrowsize=0.05291667cm 2.0,arrowlength=1.4,arrowinset=0.0]{->}(4.1500144,1.2946808)(4.2600145,1.5796809)
\psline[linecolor=colour5, linewidth=0.04, arrowsize=0.05291667cm 2.0,arrowlength=1.4,arrowinset=0.0]{->}(5.6550145,1.1496809)(5.8250146,1.4496809)
\psline[linecolor=colour5, linewidth=0.04, arrowsize=0.05291667cm 2.0,arrowlength=1.4,arrowinset=0.0]{->}(5.3300147,1.1096809)(5.5000143,1.4096808)
\psline[linecolor=colour5, linewidth=0.04, arrowsize=0.05291667cm 2.0,arrowlength=1.4,arrowinset=0.0]{->}(4.9900146,1.0946809)(5.0850143,1.4446809)
\psline[linecolor=colour5, linewidth=0.04, arrowsize=0.05291667cm 2.0,arrowlength=1.4,arrowinset=0.0]{->}(4.6100144,0.9646809)(4.7400146,1.2946808)
\psline[linecolor=colour5, linewidth=0.04, arrowsize=0.05291667cm 2.0,arrowlength=1.4,arrowinset=0.0]{->}(4.2250147,0.9796809)(4.4150143,1.3996809)
\psline[linecolor=colour5, linewidth=0.04, arrowsize=0.05291667cm 2.0,arrowlength=1.4,arrowinset=0.0]{->}(5.6050143,1.8146809)(5.7750144,2.1146808)
\psline[linecolor=colour5, linewidth=0.04, arrowsize=0.05291667cm 2.0,arrowlength=1.4,arrowinset=0.0]{->}(5.9300146,1.7846808)(6.1000147,2.0846808)
\psline[linecolor=colour5, linewidth=0.04, arrowsize=0.05291667cm 2.0,arrowlength=1.4,arrowinset=0.0]{->}(5.8950143,1.5346808)(6.0650144,1.8346809)
\psline[linecolor=colour5, linewidth=0.04, arrowsize=0.05291667cm 2.0,arrowlength=1.4,arrowinset=0.0]{->}(5.5300145,1.5096809)(5.7000146,1.8096809)
\psline[linecolor=colour5, linewidth=0.04, arrowsize=0.05291667cm 2.0,arrowlength=1.4,arrowinset=0.0]{->}(5.1500144,1.4546809)(5.2150145,1.7396809)
\psline[linecolor=colour5, linewidth=0.04, arrowsize=0.05291667cm 2.0,arrowlength=1.4,arrowinset=0.0]{->}(4.8850145,1.7996808)(5.0550146,2.099681)
\psline[linecolor=colour5, linewidth=0.04, arrowsize=0.05291667cm 2.0,arrowlength=1.4,arrowinset=0.0]{->}(4.5350146,1.7846808)(4.7050147,2.0846808)
\psline[linecolor=colour5, linewidth=0.04, arrowsize=0.05291667cm 2.0,arrowlength=1.4,arrowinset=0.0]{->}(4.2400146,1.8596809)(4.4100146,2.1596808)
\psline[linecolor=colour5, linewidth=0.04, arrowsize=0.05291667cm 2.0,arrowlength=1.4,arrowinset=0.0]{->}(0.21001449,1.0996809)(0.0700145,1.4096808)
\psline[linecolor=colour5, linewidth=0.04, arrowsize=0.05291667cm 2.0,arrowlength=1.4,arrowinset=0.0]{->}(0.5250145,1.0296808)(0.3800145,1.3896809)
\end{pspicture}
}
\caption{Finite propagation speed.}
\label{finitepropagation}
\end{figure}
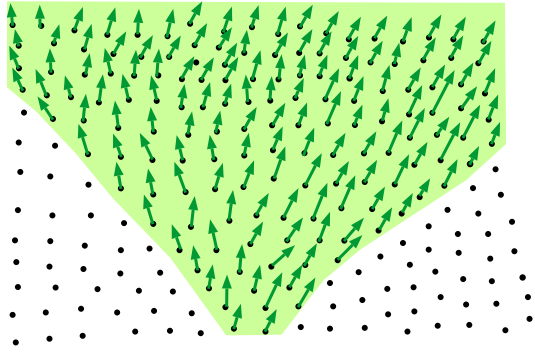

Despite this nonlocality, the linearized field equations describe a dynamics
with {\em{finite propagation speed}} (as shown schematically in Figure~\ref{finitepropagation}).
Moreover, {\em{causation}} holds in the sense that the initial value problem is well-posed.
It is important to note, however, that these statements hold only ``macroscopically'' on a scale which is much larger
than the length scale of the discretization. This is made mathematically precise in the
recent paper~\cite{linhyp}.
Before explaining in Section~\ref{sechyp} how these statements come about and what precisely we mean by ``macroscopic,'' we need to discuss a few other structures which arise for minimizers of causal variational principles.

\section{Surface Layer Integrals} \label{secosi}
\begin{figure}
%
\psscalebox{1.0 1.0} 
{
\begin{pspicture}(0,-2.172565)(7.633704,2.172565)
\definecolor{colour2}{rgb}{1.0,1.0,0.4}
\definecolor{colour5}{rgb}{0.0,0.6,0.2}
\pspolygon[linecolor=colour2, linewidth=0.04, fillstyle=solid,fillcolor=colour2](0.021402044,-1.607637)(7.611402,-1.5876371)(7.211402,2.082363)(0.27140203,2.152363)
\pspolygon[linecolor=colour5, linewidth=0.04, fillstyle=solid,fillcolor=colour5](1.091402,0.19236296)(2.781402,0.022362957)(4.8714023,0.08236296)(6.111402,0.092362955)(6.161402,-1.597637)(1.101402,-1.597637)
\rput[bl](6.554513,-0.78363705){$\Omega$}
\psdots[linecolor=black, dotsize=0.08](1.511402,1.4323629)
\psdots[linecolor=black, dotsize=0.08](2.631402,1.272363)
\psdots[linecolor=black, dotsize=0.08](3.816402,1.1773629)
\psdots[linecolor=black, dotsize=0.08](5.086402,1.3073629)
\psdots[linecolor=black, dotsize=0.08](1.501402,1.127363)
\psdots[linecolor=black, dotsize=0.08](4.146402,1.202363)
\psdots[linecolor=black, dotsize=0.08](4.651402,1.242363)
\psdots[linecolor=black, dotsize=0.08](4.501402,0.96236295)
\psdots[linecolor=black, dotsize=0.08](1.901402,0.737363)
\psdots[linecolor=black, dotsize=0.08](5.126402,1.017363)
\psdots[linecolor=black, dotsize=0.08](4.881402,1.3123629)
\psdots[linecolor=black, dotsize=0.08](4.791402,0.982363)
\psdots[linecolor=black, dotsize=0.08](1.876402,1.362363)
\psdots[linecolor=black, dotsize=0.08](2.256402,1.3073629)
\psdots[linecolor=black, dotsize=0.08](2.971402,1.242363)
\psdots[linecolor=black, dotsize=0.08](3.331402,1.207363)
\psdots[linecolor=black, dotsize=0.08](3.586402,1.1873629)
\psdots[linecolor=black, dotsize=0.08](4.421402,1.217363)
\psdots[linecolor=black, dotsize=0.08](3.2714021,0.89736295)
\psdots[linecolor=black, dotsize=0.08](3.611402,0.89236295)
\psdots[linecolor=black, dotsize=0.08](3.8964021,0.89236295)
\psdots[linecolor=black, dotsize=0.08](4.191402,0.93736297)
\psdots[linecolor=black, dotsize=0.08](1.8214021,1.0523629)
\psdots[linecolor=black, dotsize=0.08](2.141402,1.027363)
\psdots[linecolor=black, dotsize=0.08](2.5114021,0.95736295)
\psdots[linecolor=black, dotsize=0.08](2.931402,0.932363)
\psdots[linecolor=black, dotsize=0.08](1.516402,0.82236296)
\psdots[linecolor=black, dotsize=0.08](2.2714021,0.72736293)
\psdots[linecolor=black, dotsize=0.08](2.7714021,0.617363)
\psdots[linecolor=black, dotsize=0.08](3.2414021,0.56736296)
\psdots[linecolor=black, dotsize=0.08](3.701402,0.52736294)
\psdots[linecolor=black, dotsize=0.08](4.081402,0.56736296)
\psdots[linecolor=black, dotsize=0.08](4.446402,0.60236293)
\psdots[linecolor=black, dotsize=0.08](4.816402,0.65236294)
\psdots[linecolor=black, dotsize=0.08](5.166402,0.65236294)
\psdots[linecolor=black, dotsize=0.08](1.4464021,0.42236295)
\psdots[linecolor=black, dotsize=0.08](1.8264021,0.41236296)
\psdots[linecolor=black, dotsize=0.08](2.296402,0.34236297)
\psdots[linecolor=black, dotsize=0.08](2.806402,0.23236296)
\psdots[linecolor=black, dotsize=0.08](1.466402,0.032362957)
\psdots[linecolor=black, dotsize=0.08](1.866402,-0.027637042)
\psdots[linecolor=black, dotsize=0.08](2.286402,-0.06763704)
\psdots[linecolor=black, dotsize=0.08](2.806402,-0.18763705)
\psdots[linecolor=black, dotsize=0.08](3.236402,-0.26763704)
\psdots[linecolor=black, dotsize=0.08](3.666402,-0.23763704)
\psdots[linecolor=black, dotsize=0.08](3.226402,0.19236296)
\psdots[linecolor=black, dotsize=0.08](3.6264021,0.18236296)
\psdots[linecolor=black, dotsize=0.08](4.046402,0.20236295)
\psdots[linecolor=black, dotsize=0.08](4.446402,0.25236297)
\psdots[linecolor=black, dotsize=0.08](4.826402,0.32236296)
\psdots[linecolor=black, dotsize=0.08](5.2464023,0.28236297)
\psdots[linecolor=black, dotsize=0.08](4.056402,-0.19763704)
\psdots[linecolor=black, dotsize=0.08](4.436402,-0.16763705)
\psdots[linecolor=black, dotsize=0.08](4.836402,-0.057637043)
\psdots[linecolor=black, dotsize=0.08](5.256402,-0.14763704)
\psdots[linecolor=black, dotsize=0.08](1.476402,-0.41763705)
\psdots[linecolor=black, dotsize=0.08](1.9364021,-0.45763704)
\psdots[linecolor=black, dotsize=0.08](2.406402,-0.54763705)
\psdots[linecolor=black, dotsize=0.08](2.846402,-0.64763707)
\psdots[linecolor=black, dotsize=0.08](3.236402,-0.65763706)
\psdots[linecolor=black, dotsize=0.08](3.736402,-0.697637)
\psdots[linecolor=black, dotsize=0.01](3.666402,-2.167637)
\psdots[linecolor=black, dotsize=0.08](4.2464023,-0.61763704)
\psdots[linecolor=black, dotsize=0.08](4.606402,-0.54763705)
\psdots[linecolor=black, dotsize=0.08](4.966402,-0.577637)
\psdots[linecolor=black, dotsize=0.08](5.366402,-0.58763707)
\psdots[linecolor=black, dotsize=0.08](1.396402,-0.80763704)
\psdots[linecolor=black, dotsize=0.08](1.876402,-0.85763705)
\psdots[linecolor=black, dotsize=0.08](2.326402,-0.947637)
\psdots[linecolor=black, dotsize=0.08](2.806402,-1.007637)
\psdots[linecolor=black, dotsize=0.08](5.446402,-0.877637)
\psdots[linecolor=black, dotsize=0.08](5.0164022,-0.877637)
\psdots[linecolor=black, dotsize=0.08](4.526402,-0.92763704)
\psdots[linecolor=black, dotsize=0.08](4.076402,-1.007637)
\psdots[linecolor=black, dotsize=0.08](3.586402,-1.007637)
\psdots[linecolor=black, dotsize=0.08](3.186402,-0.98763704)
\psdots[linecolor=black, dotsize=0.08](1.4164021,-1.167637)
\psdots[linecolor=black, dotsize=0.08](1.856402,-1.227637)
\psdots[linecolor=black, dotsize=0.08](2.306402,-1.257637)
\psdots[linecolor=black, dotsize=0.08](2.796402,-1.3176371)
\psdots[linecolor=black, dotsize=0.08](3.326402,-1.3076371)
\psdots[linecolor=black, dotsize=0.08](3.826402,-1.277637)
\psdots[linecolor=black, dotsize=0.08](4.386402,-1.297637)
\psdots[linecolor=black, dotsize=0.08](4.806402,-1.227637)
\psdots[linecolor=black, dotsize=0.08](5.256402,-1.2076371)
\psdots[linecolor=black, dotsize=0.08](5.586402,-1.4476371)
\psdots[linecolor=black, dotsize=0.08](5.046402,-1.507637)
\psdots[linecolor=black, dotsize=0.08](4.576402,-1.547637)
\psdots[linecolor=black, dotsize=0.08](2.496402,-1.527637)
\psdots[linecolor=black, dotsize=0.08](1.9364021,-1.507637)
\psdots[linecolor=black, dotsize=0.08](1.4364021,-1.497637)
\psdots[linecolor=black, dotsize=0.08](3.046402,-1.547637)
\psdots[linecolor=black, dotsize=0.08](3.976402,-1.517637)
\psdots[linecolor=black, dotsize=0.08](1.526402,1.762363)
\psdots[linecolor=black, dotsize=0.08](1.9164021,1.742363)
\psdots[linecolor=black, dotsize=0.08](2.316402,1.6623629)
\psdots[linecolor=black, dotsize=0.08](2.706402,1.602363)
\psdots[linecolor=black, dotsize=0.08](3.076402,1.5623629)
\psdots[linecolor=black, dotsize=0.08](3.466402,1.512363)
\psdots[linecolor=black, dotsize=0.08](3.806402,1.487363)
\psdots[linecolor=black, dotsize=0.08](4.106402,1.4173629)
\psdots[linecolor=black, dotsize=0.08](4.406402,1.5373629)
\psdots[linecolor=black, dotsize=0.08](4.7664022,1.5473629)
\psdots[linecolor=black, dotsize=0.08](5.176402,1.617363)
\psdots[linecolor=black, dotsize=0.08](1.746402,1.977363)
\psdots[linecolor=black, dotsize=0.08](2.186402,1.9173629)
\psdots[linecolor=black, dotsize=0.08](2.6264021,1.977363)
\psdots[linecolor=black, dotsize=0.08](3.026402,1.857363)
\psdots[linecolor=black, dotsize=0.08](3.426402,1.837363)
\psdots[linecolor=black, dotsize=0.08](3.786402,1.717363)
\psdots[linecolor=black, dotsize=0.08](4.176402,1.897363)
\psdots[linecolor=black, dotsize=0.08](4.106402,1.6873629)
\psdots[linecolor=black, dotsize=0.08](3.826402,2.047363)
\psdots[linecolor=black, dotsize=0.08](4.546402,1.957363)
\psdots[linecolor=black, dotsize=0.08](4.896402,1.867363)
\psdots[linecolor=black, dotsize=0.08](5.186402,1.967363)
\rput[bl](5.554513,1.286363){$M \setminus \Omega$}
\end{pspicture}
}
\caption{A surface layer integral.}
\label{osi}
\end{figure}
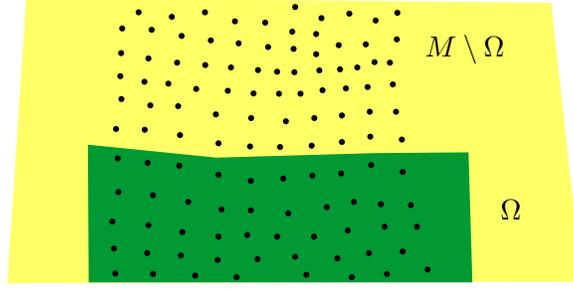

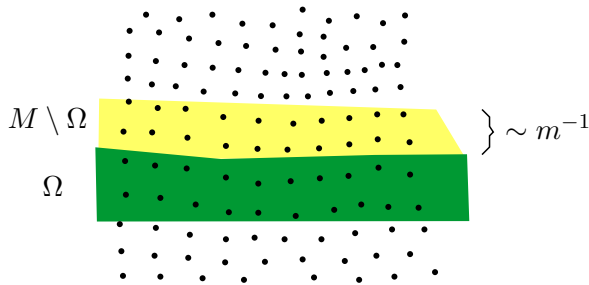
\begin{figure}
%
\psscalebox{1.0 1.0} 
{
\begin{pspicture}(0,-2.1296756)(7.62,2.1296756)
\definecolor{colour0}{rgb}{1.0,1.0,0.4}
\definecolor{colour1}{rgb}{0.0,0.6,0.2}
\pspolygon[linecolor=colour0, linewidth=0.04, fillstyle=solid,fillcolor=colour0](1.2218889,0.12525246)(6.051889,0.12525246)(5.681101,0.74049056)(1.2316464,0.8852525)
\pspolygon[linecolor=colour1, linewidth=0.04, fillstyle=solid,fillcolor=colour1](1.1818889,0.23525245)(2.831916,0.0812288)(4.8724823,0.13559009)(6.083153,0.14465031)(6.111889,-0.7045007)(1.2014159,-0.7103368)
\rput[bl](0.46,-0.39574754){$\Omega$}
\psdots[linecolor=black, dotsize=0.08](1.6018889,1.4752525)
\psdots[linecolor=black, dotsize=0.08](2.7218888,1.3152524)
\psdots[linecolor=black, dotsize=0.08](3.906889,1.2202525)
\psdots[linecolor=black, dotsize=0.08](5.176889,1.3502525)
\psdots[linecolor=black, dotsize=0.08](1.5918889,1.1702524)
\psdots[linecolor=black, dotsize=0.08](4.236889,1.2452525)
\psdots[linecolor=black, dotsize=0.08](4.741889,1.2852525)
\psdots[linecolor=black, dotsize=0.08](4.591889,1.0052525)
\psdots[linecolor=black, dotsize=0.08](1.9918889,0.78025246)
\psdots[linecolor=black, dotsize=0.08](5.216889,1.0602524)
\psdots[linecolor=black, dotsize=0.08](4.971889,1.3552525)
\psdots[linecolor=black, dotsize=0.08](4.881889,1.0252525)
\psdots[linecolor=black, dotsize=0.08](1.9668889,1.4052525)
\psdots[linecolor=black, dotsize=0.08](2.3468888,1.3502525)
\psdots[linecolor=black, dotsize=0.08](3.061889,1.2852525)
\psdots[linecolor=black, dotsize=0.08](3.4218888,1.2502525)
\psdots[linecolor=black, dotsize=0.08](3.676889,1.2302525)
\psdots[linecolor=black, dotsize=0.08](4.511889,1.2602525)
\psdots[linecolor=black, dotsize=0.08](3.361889,0.9402525)
\psdots[linecolor=black, dotsize=0.08](3.7018888,0.9352524)
\psdots[linecolor=black, dotsize=0.08](3.986889,0.9352524)
\psdots[linecolor=black, dotsize=0.08](4.281889,0.98025244)
\psdots[linecolor=black, dotsize=0.08](1.9118888,1.0952525)
\psdots[linecolor=black, dotsize=0.08](2.2318888,1.0702524)
\psdots[linecolor=black, dotsize=0.08](2.601889,1.0002525)
\psdots[linecolor=black, dotsize=0.08](3.021889,0.97525245)
\psdots[linecolor=black, dotsize=0.08](1.6068889,0.86525244)
\psdots[linecolor=black, dotsize=0.08](2.361889,0.77025247)
\psdots[linecolor=black, dotsize=0.08](2.861889,0.66025245)
\psdots[linecolor=black, dotsize=0.08](3.331889,0.61025244)
\psdots[linecolor=black, dotsize=0.08](3.791889,0.5702525)
\psdots[linecolor=black, dotsize=0.08](4.171889,0.61025244)
\psdots[linecolor=black, dotsize=0.08](4.536889,0.64525247)
\psdots[linecolor=black, dotsize=0.08](4.906889,0.6952525)
\psdots[linecolor=black, dotsize=0.08](5.256889,0.6952525)
\psdots[linecolor=black, dotsize=0.08](1.5368888,0.46525246)
\psdots[linecolor=black, dotsize=0.08](1.9168888,0.45525247)
\psdots[linecolor=black, dotsize=0.08](2.386889,0.38525245)
\psdots[linecolor=black, dotsize=0.08](2.896889,0.27525246)
\psdots[linecolor=black, dotsize=0.08](1.5568889,0.07525246)
\psdots[linecolor=black, dotsize=0.08](1.9568889,0.015252457)
\psdots[linecolor=black, dotsize=0.08](2.376889,-0.024747543)
\psdots[linecolor=black, dotsize=0.08](2.896889,-0.14474754)
\psdots[linecolor=black, dotsize=0.08](3.3268888,-0.22474754)
\psdots[linecolor=black, dotsize=0.08](3.7568889,-0.19474754)
\psdots[linecolor=black, dotsize=0.08](3.3168888,0.23525245)
\psdots[linecolor=black, dotsize=0.08](3.716889,0.22525245)
\psdots[linecolor=black, dotsize=0.08](4.136889,0.24525246)
\psdots[linecolor=black, dotsize=0.08](4.536889,0.29525244)
\psdots[linecolor=black, dotsize=0.08](4.9168887,0.36525247)
\psdots[linecolor=black, dotsize=0.08](5.336889,0.32525244)
\psdots[linecolor=black, dotsize=0.08](4.1468887,-0.15474755)
\psdots[linecolor=black, dotsize=0.08](4.526889,-0.124747545)
\psdots[linecolor=black, dotsize=0.08](4.926889,-0.014747543)
\psdots[linecolor=black, dotsize=0.08](5.346889,-0.10474754)
\psdots[linecolor=black, dotsize=0.08](1.5668889,-0.37474754)
\psdots[linecolor=black, dotsize=0.08](2.0268888,-0.41474754)
\psdots[linecolor=black, dotsize=0.08](2.4968889,-0.50474757)
\psdots[linecolor=black, dotsize=0.08](2.936889,-0.60474753)
\psdots[linecolor=black, dotsize=0.08](3.3268888,-0.6147475)
\psdots[linecolor=black, dotsize=0.08](3.8268888,-0.65474755)
\psdots[linecolor=black, dotsize=0.01](3.7568889,-2.1247475)
\psdots[linecolor=black, dotsize=0.08](4.336889,-0.57474756)
\psdots[linecolor=black, dotsize=0.08](4.696889,-0.50474757)
\psdots[linecolor=black, dotsize=0.08](5.056889,-0.53474754)
\psdots[linecolor=black, dotsize=0.08](5.4568887,-0.54474753)
\psdots[linecolor=black, dotsize=0.08](1.4868889,-0.76474756)
\psdots[linecolor=black, dotsize=0.08](1.9668889,-0.8147476)
\psdots[linecolor=black, dotsize=0.08](2.416889,-0.90474755)
\psdots[linecolor=black, dotsize=0.08](2.896889,-0.96474755)
\psdots[linecolor=black, dotsize=0.08](5.536889,-0.83474755)
\psdots[linecolor=black, dotsize=0.08](5.106889,-0.83474755)
\psdots[linecolor=black, dotsize=0.08](4.616889,-0.88474756)
\psdots[linecolor=black, dotsize=0.08](4.1668887,-0.96474755)
\psdots[linecolor=black, dotsize=0.08](3.676889,-0.96474755)
\psdots[linecolor=black, dotsize=0.08](3.2768888,-0.94474757)
\psdots[linecolor=black, dotsize=0.08](1.5068889,-1.1247475)
\psdots[linecolor=black, dotsize=0.08](1.9468889,-1.1847476)
\psdots[linecolor=black, dotsize=0.08](2.396889,-1.2147475)
\psdots[linecolor=black, dotsize=0.08](2.886889,-1.2747475)
\psdots[linecolor=black, dotsize=0.08](3.416889,-1.2647475)
\psdots[linecolor=black, dotsize=0.08](3.916889,-1.2347475)
\psdots[linecolor=black, dotsize=0.08](4.4768887,-1.2547475)
\psdots[linecolor=black, dotsize=0.08](4.8968887,-1.1847476)
\psdots[linecolor=black, dotsize=0.08](5.346889,-1.1647476)
\psdots[linecolor=black, dotsize=0.08](5.676889,-1.4047475)
\psdots[linecolor=black, dotsize=0.08](5.136889,-1.4647475)
\psdots[linecolor=black, dotsize=0.08](4.6668887,-1.5047475)
\psdots[linecolor=black, dotsize=0.08](2.5868888,-1.4847475)
\psdots[linecolor=black, dotsize=0.08](2.0268888,-1.4647475)
\psdots[linecolor=black, dotsize=0.08](1.5268888,-1.4547476)
\psdots[linecolor=black, dotsize=0.08](3.136889,-1.5047475)
\psdots[linecolor=black, dotsize=0.08](4.066889,-1.4747475)
\psdots[linecolor=black, dotsize=0.08](1.6168889,1.8052524)
\psdots[linecolor=black, dotsize=0.08](2.0068889,1.7852525)
\psdots[linecolor=black, dotsize=0.08](2.406889,1.7052524)
\psdots[linecolor=black, dotsize=0.08](2.7968888,1.6452525)
\psdots[linecolor=black, dotsize=0.08](3.166889,1.6052525)
\psdots[linecolor=black, dotsize=0.08](3.5568888,1.5552524)
\psdots[linecolor=black, dotsize=0.08](3.896889,1.5302525)
\psdots[linecolor=black, dotsize=0.08](4.196889,1.4602524)
\psdots[linecolor=black, dotsize=0.08](4.496889,1.5802524)
\psdots[linecolor=black, dotsize=0.08](4.856889,1.5902524)
\psdots[linecolor=black, dotsize=0.08](5.266889,1.6602525)
\psdots[linecolor=black, dotsize=0.08](1.8368889,2.0202525)
\psdots[linecolor=black, dotsize=0.08](2.2768888,1.9602524)
\psdots[linecolor=black, dotsize=0.08](2.716889,2.0202525)
\psdots[linecolor=black, dotsize=0.08](3.116889,1.9002525)
\psdots[linecolor=black, dotsize=0.08](3.5168889,1.8802525)
\psdots[linecolor=black, dotsize=0.08](3.876889,1.7602525)
\psdots[linecolor=black, dotsize=0.08](4.266889,1.9402524)
\psdots[linecolor=black, dotsize=0.08](4.196889,1.7302525)
\psdots[linecolor=black, dotsize=0.08](3.916889,2.0902524)
\psdots[linecolor=black, dotsize=0.08](4.636889,2.0002525)
\psdots[linecolor=black, dotsize=0.08](4.986889,1.9102525)
\psdots[linecolor=black, dotsize=0.08](5.276889,2.0102525)
\rput[bl](0.0,0.41925246){$M \setminus \Omega$}
\psline[linecolor=black, linewidth=0.02](6.301889,0.7602525)(6.401889,0.66025245)(6.401889,0.53525245)(6.451889,0.46025246)(6.401889,0.41025245)(6.401889,0.26025245)(6.301889,0.16025245)
\rput[bl](6.61,0.37425247){$\sim m^{-1}$}
\end{pspicture}
}
\caption{Localization of a surface layer integral.}
\label{osi2}
\end{figure}

In analogy to the situation for the classical Noether theorem,
symmetries of minimizers of the causal action
give rise to corresponding conservation laws.
In the resulting {\em{Noether-like theorems}}~\cite{noether},
the conserved quantities are expressed in terms of
so-called {\em{surface layer integrals}} which have the following general structure.
We choose a subset~$\Omega$ of space-time which can be thought of
as the past of an equal time hypersurface of an observer (see Figure~\ref{osi}).
We now form a double integral where one variable~$x$ is integrated over~$\Omega$
and the other variable~$y$ over the complement of~$\Omega$,
\[ \int_\Omega d\rho(x) \int_{M \setminus \Omega} d\rho(y)\: (\cdots) \L(x,y) \:. \]
Here~$(\cdots)$ stands for a an operator acting on the Lagrangian
which typically involves jet derivatives.
For specific choices of this operator, one recovers conservation laws for
charge, energy and angular momentum.
The structure of a surface integral can be understood most easily
if one keeps in mind that in the applications, the Lagrangian~$\L(x,y)$
as well as its jet derivatives are of short range. More precisely, these functions
decay on the Compton scale~$m^{-1}$. As a consequence, the relevant contribution
to the surface layer integral is obtained when both~$x$ and~$y$ are close to the boundary
of~$\Omega$. As a result, a surface layer integral can be thought of as an
integral over a ``time strip'' of size~$m^{-1}$ (see Figure~\ref{osi2}).

There are also conservation laws which are not related to symmetries of space-time.
Instead, they are a direct consequence of the linearized field equations.
Again, the corresponding conserved quantities are expressed in terms of surface layer
integrals. From the general class of conservation laws found in~\cite{osi},
the most useful surface layer integrals for the applications are
the {\em{symplectic form}}~$\sigma_\Omega$ (see~\cite[Sections~3.3 and~4.3]{jet})
\beq \label{symplectic}
\sigma_\Omega(\u,\v) = \int_\Omega d\rho(x) \int_{M \setminus \Omega} d\rho(y)\; 
\big( \nabla_{1, \u} \nabla_{2, \v} - \nabla_{1, \v} \nabla_{2, \u} \big) \, \L(x,y)
\eeq
and the {\em{surface layer inner product}}~$(.,.)_\Omega$ (see~\cite[Theorem~1.1]{osi})
\beq \label{osisp}
(\u,\v)_\Omega = \int_\Omega d\rho(x) \int_{M \setminus \Omega} d\rho(y)\; 
\big( \nabla_{1, \u} \nabla_{1, \v} - \nabla_{2, \u} \nabla_{2, \v} \big) \, \L(x,y) \:.
\eeq
Note that the surface layer inner product is symmetric in its two arguments.
In~\cite{action} it is shown by explicit computation that it is positive definite
for Dirac sea configurations in Minkowski space.
Therefore, it defines a scalar product on the real vector space of linearized solutions.

\section{Unitary Time Evolution on Fock Spaces}
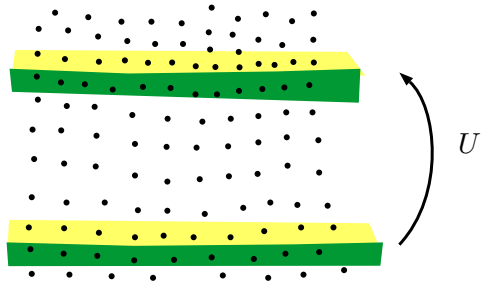
\begin{figure}[t]
%
\psscalebox{1.0 1.0} 
{
\begin{pspicture}(0,-2.1296756)(6.244617,2.1296756)
\definecolor{colour2}{rgb}{1.0,1.0,0.4}
\definecolor{colour5}{rgb}{0.0,0.6,0.2}
\pspolygon[linecolor=colour2, linewidth=0.04, fillstyle=solid,fillcolor=colour2](0.08150604,1.1352525)(4.731506,1.2052524)(4.518095,1.4804906)(0.13163178,1.5052525)
\pspolygon[linecolor=colour5, linewidth=0.04, fillstyle=solid,fillcolor=colour5](0.061506044,1.2552525)(1.6170847,1.1926593)(3.5923264,1.2200055)(4.6827703,1.2546861)(4.6636057,0.84476167)(0.0896517,0.9880643)
\pspolygon[linecolor=colour2, linewidth=0.04, fillstyle=solid,fillcolor=colour2](0.031506043,-1.1347475)(4.951506,-1.1347475)(4.808095,-0.79950947)(0.05163177,-0.75474757)
\pspolygon[linecolor=colour5, linewidth=0.04, fillstyle=solid,fillcolor=colour5](0.021506043,-1.0547476)(1.6915332,-1.1005675)(3.8120992,-1.0805492)(4.98277,-1.0551621)(4.951506,-1.3252383)(0.041032992,-1.3303368)
\psdots[linecolor=black, dotsize=0.08](0.41150606,1.4752525)
\psdots[linecolor=black, dotsize=0.08](1.571506,1.3852525)
\psdots[linecolor=black, dotsize=0.08](2.776506,1.3002524)
\psdots[linecolor=black, dotsize=0.08](4.086506,1.3602525)
\psdots[linecolor=black, dotsize=0.08](0.40150604,1.1702524)
\rput{98.70961}(4.845716,-1.5694104){\psdots[linecolor=black, dotsize=0.08](3.096506,1.2952524)}
\psdots[linecolor=black, dotsize=0.08](3.561506,1.3252524)
\psdots[linecolor=black, dotsize=0.08](3.401506,1.0052525)
\psdots[linecolor=black, dotsize=0.08](0.80150604,0.78025246)
\psdots[linecolor=black, dotsize=0.08](4.026506,1.0602524)
\psdots[linecolor=black, dotsize=0.08](3.811506,1.3652525)
\psdots[linecolor=black, dotsize=0.08](3.6915061,1.0252525)
\psdots[linecolor=black, dotsize=0.08](0.77650607,1.4052525)
\psdots[linecolor=black, dotsize=0.08](1.226506,1.3702525)
\psdots[linecolor=black, dotsize=0.08](1.8815061,1.3552525)
\psdots[linecolor=black, dotsize=0.08](2.2015061,1.3602525)
\rput{20.632116}(0.62309206,-0.80269516){\psdots[linecolor=black, dotsize=0.08](2.516506,1.3102524)}
\rput{14.352597}(0.4368384,-0.7889685){\psdots[linecolor=black, dotsize=0.08](3.351506,1.3402524)}
\psdots[linecolor=black, dotsize=0.08](2.1815062,1.0202525)
\psdots[linecolor=black, dotsize=0.08](2.511506,0.9352524)
\psdots[linecolor=black, dotsize=0.08](2.796506,0.9352524)
\psdots[linecolor=black, dotsize=0.08](3.091506,0.98025244)
\psdots[linecolor=black, dotsize=0.08](0.72150606,1.0952525)
\psdots[linecolor=black, dotsize=0.08](1.041506,1.0702524)
\psdots[linecolor=black, dotsize=0.08](1.411506,1.0002525)
\psdots[linecolor=black, dotsize=0.08](1.811506,1.0352525)
\psdots[linecolor=black, dotsize=0.08](0.41650605,0.86525244)
\psdots[linecolor=black, dotsize=0.08](1.171506,0.77025247)
\psdots[linecolor=black, dotsize=0.08](1.671506,0.66025245)
\psdots[linecolor=black, dotsize=0.08](2.141506,0.61025244)
\psdots[linecolor=black, dotsize=0.08](2.601506,0.5702525)
\psdots[linecolor=black, dotsize=0.08](2.981506,0.61025244)
\psdots[linecolor=black, dotsize=0.08](3.346506,0.64525247)
\psdots[linecolor=black, dotsize=0.08](3.716506,0.6952525)
\psdots[linecolor=black, dotsize=0.08](4.066506,0.6952525)
\psdots[linecolor=black, dotsize=0.08](0.34650603,0.46525246)
\psdots[linecolor=black, dotsize=0.08](0.72650605,0.45525247)
\psdots[linecolor=black, dotsize=0.08](1.196506,0.38525245)
\psdots[linecolor=black, dotsize=0.08](1.706506,0.27525246)
\psdots[linecolor=black, dotsize=0.08](0.36650604,0.07525246)
\psdots[linecolor=black, dotsize=0.08](0.766506,0.015252457)
\psdots[linecolor=black, dotsize=0.08](1.186506,-0.024747543)
\psdots[linecolor=black, dotsize=0.08](1.706506,-0.14474754)
\psdots[linecolor=black, dotsize=0.08](2.136506,-0.22474754)
\psdots[linecolor=black, dotsize=0.08](2.5665061,-0.19474754)
\psdots[linecolor=black, dotsize=0.08](2.126506,0.23525245)
\psdots[linecolor=black, dotsize=0.08](2.526506,0.22525245)
\psdots[linecolor=black, dotsize=0.08](2.946506,0.24525246)
\psdots[linecolor=black, dotsize=0.08](3.346506,0.29525244)
\psdots[linecolor=black, dotsize=0.08](3.726506,0.36525247)
\psdots[linecolor=black, dotsize=0.08](4.146506,0.32525244)
\psdots[linecolor=black, dotsize=0.08](2.956506,-0.15474755)
\psdots[linecolor=black, dotsize=0.08](3.3365061,-0.124747545)
\psdots[linecolor=black, dotsize=0.08](3.736506,-0.014747543)
\psdots[linecolor=black, dotsize=0.08](4.156506,-0.10474754)
\psdots[linecolor=black, dotsize=0.08](0.33650604,-0.43474755)
\psdots[linecolor=black, dotsize=0.08](0.79650605,-0.50474757)
\psdots[linecolor=black, dotsize=0.08](1.306506,-0.50474757)
\psdots[linecolor=black, dotsize=0.08](1.7465061,-0.60474753)
\psdots[linecolor=black, dotsize=0.08](2.136506,-0.6147475)
\psdots[linecolor=black, dotsize=0.08](2.636506,-0.65474755)
\psdots[linecolor=black, dotsize=0.01](2.5665061,-2.1247475)
\psdots[linecolor=black, dotsize=0.08](3.146506,-0.57474756)
\psdots[linecolor=black, dotsize=0.08](3.506506,-0.50474757)
\psdots[linecolor=black, dotsize=0.08](3.866506,-0.53474754)
\psdots[linecolor=black, dotsize=0.08](4.266506,-0.54474753)
\psdots[linecolor=black, dotsize=0.08](0.29650605,-0.83474755)
\psdots[linecolor=black, dotsize=0.08](0.766506,-0.84474754)
\psdots[linecolor=black, dotsize=0.08](1.226506,-0.90474755)
\psdots[linecolor=black, dotsize=0.08](1.706506,-0.96474755)
\psdots[linecolor=black, dotsize=0.08](4.346506,-0.83474755)
\psdots[linecolor=black, dotsize=0.08](3.916506,-0.83474755)
\psdots[linecolor=black, dotsize=0.08](3.426506,-0.88474756)
\psdots[linecolor=black, dotsize=0.08](2.976506,-0.96474755)
\psdots[linecolor=black, dotsize=0.08](2.486506,-0.96474755)
\psdots[linecolor=black, dotsize=0.08](2.0865061,-0.94474757)
\psdots[linecolor=black, dotsize=0.08](0.31650603,-1.1247475)
\psdots[linecolor=black, dotsize=0.08](0.756506,-1.1847476)
\psdots[linecolor=black, dotsize=0.08](1.206506,-1.2147475)
\psdots[linecolor=black, dotsize=0.08](1.696506,-1.2747475)
\psdots[linecolor=black, dotsize=0.08](2.226506,-1.2647475)
\psdots[linecolor=black, dotsize=0.08](2.726506,-1.2347475)
\psdots[linecolor=black, dotsize=0.08](3.286506,-1.2547475)
\psdots[linecolor=black, dotsize=0.08](3.706506,-1.1847476)
\psdots[linecolor=black, dotsize=0.08](4.156506,-1.1647476)
\psdots[linecolor=black, dotsize=0.08](4.486506,-1.4047475)
\psdots[linecolor=black, dotsize=0.08](3.946506,-1.4647475)
\psdots[linecolor=black, dotsize=0.08](3.476506,-1.5047475)
\psdots[linecolor=black, dotsize=0.08](1.3965061,-1.4847475)
\psdots[linecolor=black, dotsize=0.08](0.83650607,-1.4647475)
\psdots[linecolor=black, dotsize=0.08](0.33650604,-1.4547476)
\psdots[linecolor=black, dotsize=0.08](1.946506,-1.5047475)
\psdots[linecolor=black, dotsize=0.08](2.876506,-1.4747475)
\psdots[linecolor=black, dotsize=0.08](0.42650604,1.8052524)
\psdots[linecolor=black, dotsize=0.08](0.816506,1.7852525)
\psdots[linecolor=black, dotsize=0.08](1.216506,1.7052524)
\psdots[linecolor=black, dotsize=0.08](1.606506,1.6452525)
\psdots[linecolor=black, dotsize=0.08](1.976506,1.6052525)
\psdots[linecolor=black, dotsize=0.08](2.346506,1.6052525)
\psdots[linecolor=black, dotsize=0.08](2.706506,1.5302525)
\psdots[linecolor=black, dotsize=0.08](3.0865061,1.5002525)
\psdots[linecolor=black, dotsize=0.08](3.346506,1.6602525)
\psdots[linecolor=black, dotsize=0.08](3.666506,1.5902524)
\psdots[linecolor=black, dotsize=0.08](4.076506,1.6602525)
\psdots[linecolor=black, dotsize=0.08](0.6465061,2.0202525)
\psdots[linecolor=black, dotsize=0.08](1.086506,1.9602524)
\psdots[linecolor=black, dotsize=0.08](1.5265061,1.9002525)
\psdots[linecolor=black, dotsize=0.08](1.926506,1.9002525)
\psdots[linecolor=black, dotsize=0.08](2.3265061,1.8802525)
\psdots[linecolor=black, dotsize=0.08](2.686506,1.7602525)
\psdots[linecolor=black, dotsize=0.08](3.0765061,1.9402524)
\psdots[linecolor=black, dotsize=0.08](3.006506,1.7302525)
\psdots[linecolor=black, dotsize=0.08](2.726506,2.0902524)
\psdots[linecolor=black, dotsize=0.08](3.446506,2.0002525)
\psdots[linecolor=black, dotsize=0.08](3.796506,1.9102525)
\psdots[linecolor=black, dotsize=0.08](4.086506,2.0102525)
\rput[bl](6.004617,0.14925246){$U$}
\psbezier[linecolor=black, linewidth=0.04, arrowsize=0.05291667cm 2.0,arrowlength=1.4,arrowinset=0.0]{->}(5.221506,-1.0647476)(5.841506,-0.3980387)(5.721506,0.77719337)(5.221506,1.2252524566650391)
\end{pspicture}
}
\caption{Unitary time evolution.}
\label{unitary}
\end{figure}

Based on the above surface layer integrals, the nonlinear dynamics as described
by the causal action principle can be rewritten in terms of a unitary time evolution
on Fock spaces, as we now outline (for details see~\cite{fockbosonic}):
Assuming that the surface layer inner product~\eqref{osisp} is positive definite and
combining it with the symplectic form~\eqref{symplectic},
one can endow the space of linearized solutions with the structure of a complex Hilbert space,
denoted by~$(\h, \la .|. \ra)$. The scalar product on~$\h$ has the property that its real part
coincides with the scalar product~$(.,.)_\Omega$. 
In the application of Dirac sea configurations in Minkowski space~\cite{action},
one sees that this scalar product agrees with the usual scalar product used
in quantum theory. In particular, it is compatible with the probabilistic interpretation.
The conservation laws for the surface layer integrals imply that the linearized fields preserve
the complex structure and the scalar product. In other words, the linearized dynamics is described by a unitary
operator~$U$ on~$\h$ (see Figure~\ref{unitary}).

In order to describe the interacting dynamics, one makes essential use of the
perturbation theory developed in~\cite{perturb}, which makes it possible to express
the dynamics of the physical system as described by the causal action principle
in terms of a nonlinear time evolution on jet spaces, which can be expanded in a linear,
a quadratic, a cubic contribution, and so on. Rewriting the resulting multilinear operators as
linear operators on the symmetric tensor product, one obtains a {\em{linear dynamics
on bosonic Fock spaces}} (for details see~\cite{fockbosonic}).
Again making use of conservation laws for surface layer integrals,
this dynamics can be written with a norm-preserving complex-linear
time evolution operator acting on the bosonic Fock space tensored with its dual.
A {\em{unitary time evolution}} on the bosonic Fock space is obtained in
the so-called holomorphic approximation.
In~\cite{fockbosonic} the error of this approximation is quantified, and it is argued
that in the physical applications in mind, this error is very small.

\section{Hyperbolicity, Causation and Finite Propagation Speed} \label{sechyp}
Having surface layer integrals to our disposal, we
can finally explain in which sense causation holds and
what we mean by finite propagation speed
(see the end of Section~\ref{seclin} and Figure~\ref{finitepropagation}).
The fact that certain surface layer integrals are conserved for solutions
of the linearized field equations can also be used for proving existence
of solutions and for analyzing their properties.
Generally speaking, one works with surface layer integrals which are
positive. These positivity properties are subsumed in so-called
{\em{hyperbolicity conditions}}.
Then one derives {\em{energy estimates}}, which basically
show that the surface layer integrals cannot grow too fast in time.
Once these energy estimates have been established, one can mimic the procedure
for hyperbolic partial differential equations to prove existence and uniqueness
of solutions of the initial value problem. These solutions respect {\em{causation}}
in the sense that the present has an influence only on the future.
Moreover, one gets {\em{finite propagation speed}}.
But since in the surface layer integrals the jets are integrated over regions of space-time
of the size of the Compton scale, the energy methods 
give us information on the jet dynamics only when
taking ``averages'' over macroscopic space-time regions.
This is what at the end of Section~\ref{secjet} we meant by the statement
that causation and finite propagation speed hold ``only macroscopically.'' For precise
mathematical statements we refer to~\cite{linhyp}.

Summarizing from a conceptual point of view, in a causal fermion system
there is no strict causation. Nevertheless, the usual notions of causality, future and past are well-defined
on scales larger than the Compton length.
Moreover, on this macroscopic scale, the dynamics is compatible with these
notions, and linearized fields propagate with finite speed.

\Thanks {{\em{Acknowledgments:}}
I would like to thank the organizers of the International Workshop
``DICE2018: Spacetime - Matter - Quantum Mechanics''
for the kind invitation to this inspiring conference.
I am grateful to Maximilian Jokel and Niky Kamran for helpful comments on the manuscript.

\providecommand{\bysame}{\leavevmode\hbox to3em{\hrulefill}\thinspace}
\providecommand{\MR}{\relax\ifhmode\unskip\space\fi MR }
\providecommand{\MRhref}[2]{%
  \href{http://www.ams.org/mathscinet-getitem?mr=#1}{#2}
}
\providecommand{\href}[2]{#2}

\end{document}